\documentclass[sigconf]{acmart}

\makeatletter
\def\@ACM@checkaffil{%
    \if@ACM@instpresent\else
    \ClassWarningNoLine{\@classname}{No institution present for an affiliation}%
    \fi
    \if@ACM@citypresent\else
    \ClassWarningNoLine{\@classname}{No city present for an affiliation}%
    \fi
    \if@ACM@countrypresent\else
        \ClassWarningNoLine{\@classname}{No country present for an affiliation}%
    \fi
}
\makeatother

\copyrightyear{2024}
\acmYear{2024}
\setcopyright{rightsretained}
\acmConference[CCS '24]{Proceedings of the 2024 ACM SIGSAC Conference on Computer and Communications Security}{October 14--18, 2024}{Salt Lake City, UT, USA}
\acmBooktitle{Proceedings of the 2024 ACM SIGSAC Conference on Computer and Communications Security (CCS '24), October 14--18, 2024, Salt Lake City, UT, USA}
\acmDOI{10.1145/3658644.3670336}
\acmISBN{979-8-4007-0636-3/24/10}

\makeatletter
\gdef\@copyrightpermission{
    \begin{minipage}{0.2\columnwidth}
        \href{ https://creativecommons.org/licenses/by/4.0/}{\includegraphics[width=0.90\textwidth]{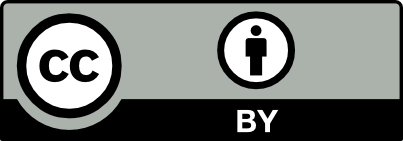}}
    \end{minipage}\hfill
    \begin{minipage}{0.8\columnwidth}
        \href{ https://creativecommons.org/licenses/by/4.0/}{This work is licensed under a Creative Commons Attribution International 4.0 License.}
    \end{minipage}
    \vspace{0pt}
}
\makeatother


\usepackage{booktabs} %
\usepackage{amsmath}
\usepackage{multirow}
\usepackage{url}
\usepackage{amsfonts}
\usepackage{balance}
\usepackage{enumitem}
\usepackage{caption}
\usepackage{tablefootnote}
\usepackage{cleveref}
\usepackage{pifont}%

\newcommand{\twolinecell}[2][r]{%
  \begin{tabular}[#1]{@{}c@{}}#2\end{tabular}}
\newcommand{\bftwolinecell}[1]{\textbf{\twolinecell{#1}}}

\usepackage{multirow}
\usepackage{makecell}
\usepackage{subcaption}
\usepackage{xspace}
\usepackage{xcolor}
\usepackage{pifont}
\usepackage{graphicx}
\usepackage{tabularx}
\usepackage{adjustbox}
\usepackage{graphicx}
\usepackage{float}

\newcommand*{\eg}{e.g.,\@\xspace}
\newcommand*{\ie}{i.e.,\@\xspace}
\newcommand*{\et}{et al.\@\xspace}

\usepackage{xstring}
\newcommand{\PP}[1]{
  \vspace{2px}
  \noindent{\bf \IfEndWith{#1}{.}{#1}{#1.}}
}

\definecolor{Green}{HTML}{00A64F}
\definecolor{Red}{HTML}{ED1B23}

\newcommand{\cmark}{{\color{Green}\ding{51}\@\xspace}}%
\newcommand{\xmark}{{\color{Red}\ding{55}\@\xspace}}%

\newcommand{\overallChangesIP}{42\%\xspace}
\newcommand{\overallChangesAS}{51\%\xspace}

\newcommand{\changesRangeHigh}{100\%\xspace}
\newcommand{\changesRangeLow}{$<$1\%\xspace}

\newcommand{\changesRangeHighHTTP}{99\%\xspace}
\newcommand{\changesRangeLowHTTP}{1\%\xspace}

\newcommand{\changesRangeHighHTTPS}{100\%\xspace}
\newcommand{\changesRangeLowHTTPS}{$<$1\%\xspace}

\newcommand{\changesRangeHighDNS}{65\%\xspace}
\newcommand{\changesRangeLowDNS}{5\%\xspace}

\newcommand{\exampleCountry}{Iran\xspace}
\newcommand{\examplehttpChange}{2\%\xspace}
\newcommand{\examplehttpsChange}{9\%\xspace}
\newcommand{\examplednsChange}{8\%\xspace}

\newcommand{\medianAffectedSparamHigh}{93\%\xspace}
\newcommand{\medianAffectedSparamLow}{1\%\xspace}

\newcommand{\medianAffectedHighWhere}{KW (\https)\xspace}
\newcommand{\medianAffectedLowWhere}{CN (\http)\xspace}

\newcommand{\maxVariation}{2x\xspace}
\newcommand{\compImpact}{35\%\xspace}

\newcommand{\rqone}{What is the path diversity of censorship measurements?\xspace}

\newcommand{\rqtwo}{What is the impact of \ecmp routing on censorship measurement results?\xspace}

\newcommand{\rqthree}{Why and how much do different packet parameters influence censorship measurement?\xspace}

\newcommand{\rqfour}{What are the underlying network structures that cause changes in censorship?\xspace}

\newcommand{\rqfive}{How do these results contextualize with specific prior works?\xspace}

\newcommand{\rqnum}{five\xspace}

\newcommand{\tcp}{\texttt{TCP}\xspace}
\newcommand{\udp}{\texttt{UDP}\xspace}
\newcommand{\icmp}{\texttt{ICMP}\xspace}
\newcommand{\http}{\texttt{HTTP}\xspace}
\newcommand{\https}{\texttt{HTTPS}\xspace}
\newcommand{\dns}{\texttt{DNS}\xspace}
\newcommand{\ttl}{\texttt{TTL}\xspace}
\newcommand{\ttls}{\texttt{TTL}s\xspace}
\newcommand{\rst}{\texttt{RST}\xspace}
\newcommand{\syn}{\texttt{SYN}\xspace}
\newcommand{\rstnospace}{\texttt{RST}}
\newcommand{\rsts}{\texttt{RST}s\xspace}

\newcommand{\ECMP}{Equal-cost Multi-path (ECMP)\xspace}
\newcommand{\ecmp}{ECMP\xspace}
\newcommand{\flowid}{Flow-ID\xspace}

\newcommand{\typeone}{\textbf{Type 1}: \ecmp routing (Inter-AS or
  Intra-AS) exercising possibly failed/misconfigured censoring nodes\xspace}

\newcommand{\typetwo}{\textbf{Type 2}: \ecmp routing through geographically
  different regions with different censorship behavior\xspace}

\newcommand{\typethree}{\textbf{Type 3}: \ecmp routing \emph{around}
  censorship\xspace}

\newcommand{\typefour}{\textbf{Type 4}: Behavior that is unknown or cannot be
  directly attributed to variation in path\xspace}

\newcommand{\numtypes}{three\xspace}

\begin{document}
\title{Understanding Routing-Induced Censorship Changes Globally}

\newcommand{\name}{\emph{Monocle}\xspace}

\author{Abhishek Bhaskar}
\email{abhaskar@gatech.edu}
\affiliation{%
  \institution{Georgia Institute of Technology}}
\author{Paul Pearce}
\email{pearce@gatech.edu}
\affiliation{%
  \institution{Georgia Institute of Technology}}

\begin{abstract}

Internet censorship is pervasive, with significant effort
dedicated to understanding what is censored, and where.  Prior censorship
measurements however have identified significant inconsistencies in their
results; experiments show unexplained non-deterministic behaviors thought to be
caused by censor load, end-host geographic diversity, or incomplete 
censorship---inconsistencies which impede reliable, repeatable and
correct understanding of global censorship.  
In this work we investigate the extent to which \ECMP routing is the cause for
these inconsistencies, developing methods to measure and compensate for them.

We find that \ecmp routing significantly changes observed censorship across
protocols, censor mechanisms, and in 17 countries. We identify that previously
observed non-determinism or regional variations are attributable to
measurements between fixed end-hosts taking different routes based on \flowid;
\ie choice of intra-subnet source IP or ephemeral source port leads to differences in
observed censorship.  To achieve this we develop new route-stable censorship
measurement methods that allow consistent measurement of \dns, \http, and
\https censorship.  We find \ecmp routing yields censorship changes across
\overallChangesIP~of IPs and \overallChangesAS~of ASes, but that impact is not
uniform.  We develop an application-level traceroute tool
to construct network paths using \emph{specific} censored packets,
leading us to identify numerous causes of the behavior, ranging from likely
failed infrastructure, to routes to the same end-host taking geographically
diverse paths which experience differences in censorship
\emph{en-route}. 
Finally, we explore our results in the context of prior global measurement studies, exploring first
the applicability of our findings to prior observed variations, and
then demonstrating how
specific experiments from two studies could be impacted by, and specific results are
explainable by, \ecmp routing. Our work points to methods for improving
future studies, reducing inconsistencies and increasing repeatability.

\vspace{-0.3em}
\end{abstract}

\maketitle

\section{Introduction}
\label{sec:intro}
Internet censorship impacts the lives of 72\% of people~\cite{freedomhousestats}, with governments and ISPs using sophistical
in-network capabilities to manipulate and disrupt
DNS~\cite{iris,satellite,censoredplanet},
HTTP~\cite{augur,ensafi14pam,censoredplanet}, and
HTTPS~\cite{raman2020measuring, censoredplanet, iclab}.  The
growing prevalence of Internet censorship has given rise to significant
measurement efforts focused on understanding the scale and scope of censorship
globally~\cite{iclab,censoredplanet,augur,iris}. 

The challenge of obtaining globally distributed hosts has made \emph{outside-in}
measurement~\cite{breadcrumb} an appealing alternative method of understanding
censorship.  Outside-in measurement leverages the symmetric nature of many
countries' censorship infrastructure to send measurements \emph{to} a vantage
point in a censored area (instead of originating from) and then observing any
actions taken against that flow. Outside-in measurements have demonstrated
censorship globally, across \dns manipulation~\cite{iris}, packet
drops~\cite{ensafi14pam, aryan2013internet}, \rst
injection~\cite{augur,ensafi14pam,quack}, and block-page
injection~\cite{raman2020measuring}. 

At the same time, prior censorship studies found inconsistency in
\dns, \http, and \https measurements, globally~\cite{iris,
  raman2020measuring, wang2017your, crandall2007conceptdoppler,
  weinberg2021chinese, anon2014towards, wright2014regional,
  xu2011internet, nisar2018incentivizing, Gill2015a, winters12foci,
  Ensafi2015a}. These appear in the form of non-uniform censorship
across a country/ISP~\cite{iris, wright2014regional, xu2011internet,
  nisar2018incentivizing, katira2023censorwatch, india2018yadav,
  Gill2015a}, variations in results over
time~\cite{raman2020measuring}, and \texttt{RST} injection not
observed for certain experiments~\cite{wang2017your}.  Inconsistency
makes understanding censorship and reproducing results challenging,
impacting our ability to develop technical and policy interventions.

Integral to the notion of outside-in censorship measurement is the idea that
measurements \emph{to} a specific vantage point will traverse at least some of
same set of the network infrastructure performing censorship as if \emph{from} a
vantage. Deeply embedded in that assumption is the concept of \ECMP routing.
Routers on the Internet use various fields of a packet to construct a flow
identifier (\flowid), and use that \flowid to assign the packet to a flow for
load balancing~\cite{paris_traceroute}.  \flowid is influenced by
``ephemeral'' fields such as source port, thus communication between a single
source IP and destination IP/port may take numerous possible routes through the
network~\cite{dminer}.  While \ecmp routing is well understood~\cite{dminer,
paris_traceroute},  
the extent to which \ecmp routing influences censorship measurement
globally across protocols is unknown.

Our work seeks to understand the extent of \ecmp routing on outside-in
censorship measurement across protocols, mechanisms, and countries,
with an eye toward \emph{why} changes in route impact observed
censorship.  Prior work exploring China's Great Firewall's
(GFW)~\cite{breadcrumb} DNS injection identified that some source
parameters resulted in variations in injected DNS censorship. Their
study is limited; they explore only China, and only \dns censorship;
both of which are known to exhibit unique behaviors among the worlds'
censors~\cite{hoang2021great}. Given such unique characteristics, it
is unclear if \ecmp-induced censorship measurement variations are an
artifact of the GFW, or if such phenomena generalize across countries
and disparate censorship deployments and protocols. It also remains
unclear \emph{why} such variations exist, and their impact on
measurement studies.

We ask: Does \ecmp routing influence outside-in censorship measurement globally, is it a
generalized phenomenon of censorship infrastructure, has it impacted prior
studies, and why?  This problem is challenging as prior
tools do not allow control of the parameters that influence route,
and prior traceroute tools either produce stable-routes of packets that are not
of interest (\eg ICMP ECHO), or produce unstable routes of application level
protocols (\eg HTTP packets).  To these ends we develop \name, a new
route-stable censorship measurement and traceroute platform able to understand
\ecmp-induced censorship changes not only across \dns, but also \http and
\https.  \name expands prior DNS tooling~\cite{breadcrumb} while also
developing new methods to measure and traceroute \rst packet injection,
packet drops, and censor blockpages, across protocols. 

We conduct a global study of 21 countries, 3
network protocols, and 4 censorship mechanisms, aimed at quantifying the effect
of \ecmp routing on both current remote censorship measurement as well as on
prior studies.  \textbf{We find that \ecmp routing
has significant impact on outside-in censorship measurement across countries, affecting 17
of 21 countries, as well as all types of protocols and mechanisms explored.} Our results
illustrate a complex entanglement of end-to-end censor activity with low-level
network behaviors that were previously thought unrelated and not considered.
We find \overallChangesIP of IPs and \overallChangesAS of ASes
show \ecmp-induced changes in measured censorship, with that impact unevenly
spread across 17 countries. 
We also find that some source IP and ephemeral
source port combinations detect up to \maxVariation more censorship than
others, between the same end-points.

We also utilize \name to conduct censorship-and-path aware traceroutes
of \emph{specific} censored packets, enabling us to reconstruct
network graphs, and explain \emph{why} variations exist. We find a
diverse set of explanations, ranging from routing within ASes sending
some packets to potentially failed infrastructure, to routes to a
single end-host spanning geographic regions with diverse censorship
\emph{on-path}.  We also explore the different forms of
observed variation in prior work, contextualizing when \ecmp routing is potentially applicable. Finally, we compare our results to
2 prior outside-in studies, showing that previously observed
non-determinism~\cite{iris} is explainable by source-parameter
selection, and a reproduction of the selection method of a prior
censorship study~\cite{raman2020measuring} using our experiments shows
as many as \compImpact end-points could experience \ecmp-induced
variations.

\PP{Contributions} Our contributions include:
\begin{itemize}[leftmargin=1.5em,nosep]

    \item Designing and deploying \name, a platform able to quantify
    the effects of \ecmp routing on censorship measurement across \dns, \http,
    and \https protocols, and across  \dns injection, \rst injection, packet
    drops, and blockpage censorship methods.  
    
    \item Finding that 17 of 21 countries explored with
    externally measurable censorship show censorship differences by
    varying intra-subnet source IP and/or source port, with the
    impact ranging from \changesRangeHigh to \changesRangeLow of
    destinations showing differences. 
    We also find all types of protocols and censor methods impacted.

    \item Finding that \ecmp-induced differences are due to a diverse set of
    properties, ranging from route differences within ASes performing
    censorship likely having failed infrastructure on some paths, 
    to routes to the same host having geographic
    diversity, and that geography demonstrating non-uniform
    censorship.

    \item Finding that previously observed
      non-determinism in a prior study~\cite{iris} is explainable by \ecmp routing, and
      replication of the selection method of a prior study~\cite{raman2020measuring} that experienced non-determinism
    shows up to \compImpact of its
      hosts could be impacted.

\end{itemize}

\noindent 
We thus argue \ecmp routing must be taken into account when measuring
in-network phenomena, and that measurement knowledgeable of
these properties can remain a valuable research method.

\section{Related Work}
\label{sec:related}

\textbf{Censorship measurement} has evolved in the past
decade to understand how censorship works,
what is censored, and how censorship changes over time. In order to
comprehensively measure censorship, studies built techniques to
perform measurements on various protocols: \dns~\cite{Farnan2016a,
  anon2014towards, Lowe2007a, iris, niaki2020triplet, satellite},
\http ~\cite{quack, weaver2009detecting, Ensafi2015a, nabi2013anatomy,
clayton2006ignoring}, \https~\cite{bock2021even, chai2019importance,
  bock2020exposing, raman2020measuring, gatlan2019south},
\texttt{HTTP/3}~\cite{elmenhorst2021web}, and \texttt{echo}~\cite{quack,
  raman2020measuring}.

Most state-sponsored censorship is deployed
at the ISP or network backbone~\cite{imc14-censorship-response,
zuckerman2010intermediary, xu2011internet} in the form of network middle-boxes
that intercept packets and perform actions based on them; such behavior affords
measurement, whereby sensitive packets are sent across the middle-box, and behaviors
are studied. 
Measurement typically
takes two forms: 1) Outside-In (Remote) measurement, where packets are sent
from \textit{outside} a country being measured towards points
\textit{inside} the country~\cite{iris, quack, raman2020measuring,
censoredplanet, ramesh2020decentralized, hoang2021great, augur, bock2021even,
niaki2020triplet, NiakiMFMGW21, aryan2013internet, xu2011internet, Farnan2016a,
weinberg2021chinese, Ensafi2015a}, or 2) Inside-Out measurement, where
measurements use observation points \textit{inside} the country from
volunteers, VPN servers, etc, to send sensitive censorship triggering
packets~\cite{iclab, hoang2019measuring}.

Both measurement types have trade-offs in 
deployability, ethical considerations, and scalability. Several global
censorship measurements employ remote measurement techniques as it
eliminates the need to have volunteers inside all the countries being
measured~\cite{iris, quack, raman2020measuring,
  censoredplanet, ramesh2020decentralized, ramesh2023network,breadcrumb,
  NiakiMFMGW21,nourin2023measuring, raman2020investigating}.  While these systems
are scalable and reduce the need for volunteers, they in-turn can only measure
censorship that is symmetric.
Measurements are generally
conducted at the \dns, \http or \https layer with countries potentially performing
censorship at any of these layers. At the \http layer, studies generally use
the \texttt{Host: } header~\cite{quack, raman2020measuring, ensafi14pam} to include a sensitive payload, and at the
\https layer a sensitive payload is encoded in the \texttt{SNI:}
field~\cite{chai2019importance, bock2020exposing} of the
\https\texttt{client-hello}, that triggers various
forms of censorship like \rst-injection, blockpages, packet drops
etc. At the \dns layer the sensitive payload (in the form of a domain) is issued as a
\dns \texttt{A?} query, that elicits \dns injection that is then used to study
censorship~\cite{iris, niaki2020triplet, hoang2021great, Farnan2016a,
weinberg2021chinese}. \name studies these protocols globally in
the context of \ecmp routing.

\PP{\ECMP Routing} Load Balancing is
widespread on the Internet. Augustin~\et~\cite{paris_traceroute}
explored multi-path routing in traceroute measurement, and
subsequently~\cite{augustin2010measuring} found that close to 72\% of the
(source, destination) pairs experienced some form of
load balancing. Recently Vermeulen~\et~\cite{dminer} found 18\% of
\ecmp routing path divergences span multiple ASes. Routers use
different components of Internet Layer (\texttt{IP}) and Transport Layer (\tcp and \udp)
like the source and destination IPs, source and destination ports,
etc, to perform these routing decisions. Routers are also known to use
various bits of the source/destination IP and port to make decisions on which up-link
to send packets~\cite{vmware_load_balancing, cisco_2007}. Thus any IP
measurements are subject to load balancing. 

\PP{Variance in Censorship Results}
\label{subsec:prior_variance}
Prior censorship studies have noted country/ISP level inconsistency in
DNS, HTTP, and HTTPS measurements~\cite{iris, raman2020measuring,
  wang2017your, crandall2007conceptdoppler, weinberg2021chinese,
  anon2014towards, wright2014regional, xu2011internet,
  katira2023censorwatch, india2018yadav, Gill2015a, winters12foci,
  Ensafi2015a}. Pearce~\et~\cite{iris} observed differences in \dns
manipulation across resolvers \emph{within} a
country. Raman~\et~\cite{raman2020measuring} noticed sporadic
blockpage injection within ISPs and
organizations. Wang~\et~\cite{wang2017your} noticed that for a small
percent of their experiments, they were successful in bypassing the
GFW without any evasion
strategy. Crandall~\et~\cite{crandall2007conceptdoppler} in as early
as 2007 observed that on 28.3\% of destinations (in China) they
observed no filtering, and even on the paths they \emph{did},
filtering appeared to be volatile during high-load periods of the
day. Rambert~\et~\cite{weinberg2021chinese} noticed differing levels
of censorship depending on the source and destination of the probes
(independent of geography). Anonymous~\cite{anon2014towards} showed
the presence of different injecting interfaces with changing source
IPs. Wright~\et~\cite{wright2014regional} and
Xu~\et~\cite{xu2011internet} both highlight geographic variation in
censorship implementation across the country of China, with different
provincial ISPs performing their own filtering (in addition to
filtering at the border). Aryan~\et~\cite{aryan2013internet} speculate
that individual ISPs can potentially implement their own blocking
mechanisms in addition to centralized
blocking (in Iran). Nisar~\et~\cite{nisar2018incentivizing}
showed differing blocking implementation by ISPs and differences even
within ISPs in Pakistan. Both Yadav~\et~\cite{india2018yadav} and
Katira~\et~\cite{katira2023censorwatch} observed differences in
censorship (in terms of domains censored) across different ISPs tested
in India. Gill~\et~\cite{Gill2015a} observed changes in 
measured censorship: across time, ISPs (within the same country), and
even URLs within the same ISP in several
countries. Winter~\et~\cite{winters12foci} found that certain Tor
relays remained reachable from VPSes in China even after several days
of the first Tor connection request from the VPS while a majority of
them were blocked. Ensafi~\et~\cite{Ensafi2015a} when studying China's
GFW behavior with respect to Tor found that GFW's failure (to block
Tor) were both persistent with routes \emph{and} sporadic.

While country and ISP-level variation in censorship
has been globally observed, there are numerous suspected caused of such variation.
These causes include (but are not limited to): geographical differences in
blocking~\cite{iris, wright2014regional, xu2011internet}, constantly
changing blocking methods~\cite{raman2020measuring}, differing ISP
implementations~\cite{nisar2018incentivizing, Gill2015a,
  katira2023censorwatch,india2018yadav}, network load (\eg time-of-day) on censoring
devices~\cite{crandall2007conceptdoppler, Ensafi2015a}, and selection of
source/destinations~\cite{weinberg2021chinese}. 
We stress that while
the goal of this work is to understand the
influence of \ecmp routing on censorship variation, 
we do not believe that all previously observed censorship variation is \ecmp-induced.

Bhaskar~\et~\cite{breadcrumb} first explored the role of routing on
the Chinese DNS censorship system.  Their study found that varying
source IP and port has an impact on the path taken by packets and
subsequently influenced the measurement of China's Great Firewall
(GFW). What remains unclear from prior work is if the observed
behaviors are an isolated artifact of the GFW and DNS injection, or a
broader behavior across the Internet and censorship measurement
at-scale.
Our work seeks to expand and generalize this prior work by systematically
exploring the wider effects of \ecmp routing on censorship globally, across both countries
and various protocols. We seek to quantify \emph{why} this behavior
exists pervasively across the Internet, occurring in disparate countries that
lack coordination, consistent network topologies, or common technical measures.

\section{Method}
\label{sec:Method}
\label{sec:method}

\begin{figure*}[]
  \centering
  \includegraphics[width=0.835\textwidth]{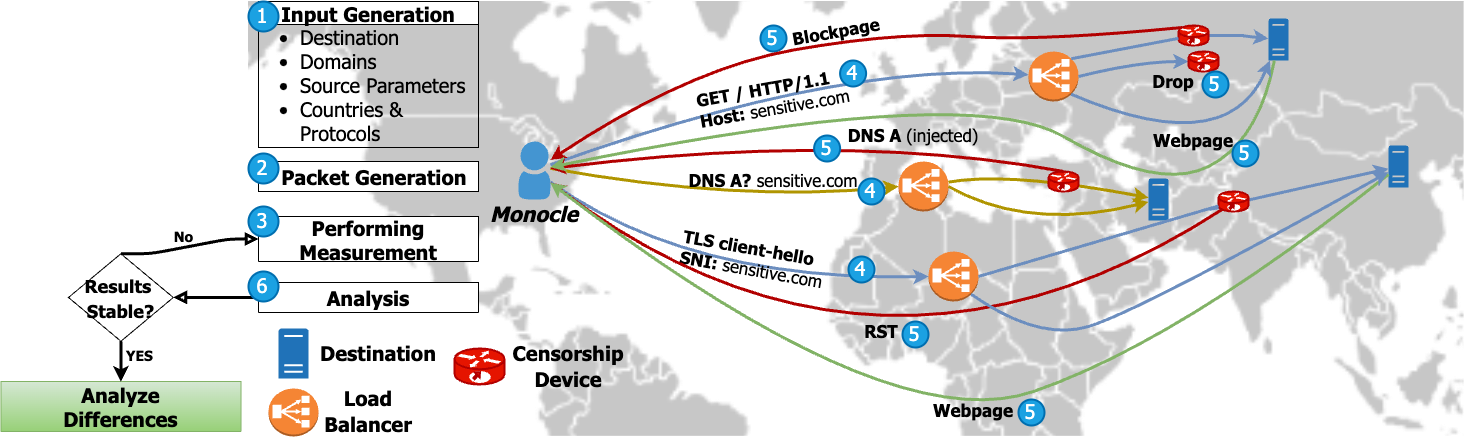}
  \caption{
    \textbf{\name}, our system to
    understand impact of routing on censorship
    measurement across countries and protocols.
  }
  \label{fig:system}
\end{figure*}

We seek to understand the impact of ephemeral parameters such as source IP (with a subnet) and source port---and
hence \flowid---on censorship measurement globally~\cite{iris, augur, quack,
raman2020measuring, censoredplanet}. \emph{Exhaustively} identifying censorship
is neither ethical nor our objective, but rather our goal is replicating
established methods from prior studies and understanding whether outside-in
censorship measurement methods are affected by \ecmp routing.

We develop \name (Figure~\ref{fig:system}) to explore 3 commonly censored protocols~\cite{gillcharacterizing}:
\dns, \http, and \https.  Across these protocols \name looks at multiple known censorship
techniques: \dns manipulation~\cite{iris}, \rst packet
injection~\cite{augur}, 
packet drops~\cite{ensafi14pam, aryan2013internet}, and
blockpages~\cite{raman2020measuring,jones14imc}. \name extends measurement methods from prior studies
to create route-stable censorship measurements, enabling us to 
measure the impact of packet parameters on censorship.

\subsection{Approach}
\label{subsec:approach}

Censorship measurement involves sending sensitive payloads between hosts that trigger a
censor, and \emph{inferring} censorship from responses (or lack thereof). \name implements
outside-in censorship measurements by sending route-stable sensitive \tcp and
\udp packets, varying source IP across a single /24 network as well as source ports, and
observing responses.
 
\PP{TCP-based censorship.} \tcp censorship works
by identifying \emph{connections} to censor, then leveraging termination or hijacking
to disrupt communications.
Censors use
\texttt{Host:} header in \http or the \texttt{SNI} field (as part of
the \texttt{client-hello}) in \https connections to perform
censorship, both of which are well studied ~\cite{quack,
  weaver2009detecting, Ensafi2015a, nabi2013anatomy,
  clayton2006ignoring, bock2021even, chai2019importance,
  bock2020exposing, raman2020measuring, gatlan2019south}. 
To disrupt communications, \tcp censorship techniques include \rst injection,
packet drops, and blockpages~\cite{censorshipglobaltechniques}.

\name establishes \tcp connections with
destination IPs inside a censored country from a vantage point outside the
country and sends a packet with keywords---either embedded in the \texttt{Host: } header of a \http
\texttt{GET} request or \texttt{SNI} field of a \https \texttt{client-hello}
request---and recording the responses to measure censorship.  \textbf{A key aspect of
\name not seen in prior work is controlling all aspects of generated
packets that are known to change \flowid, thus allowing us to perform
route-stable censorship measurements.}  We perform trials with control and
sensitive domains to establish measurement reliability. We test destinations that have ports 80 for
\http and 443 for \https, open. We identify candidate IPs via
Censys~\cite{censys} and then confirm their status.

To identify \rst based censorship we look for the presence of \rsts in our
response packets in addition to \emph{no} \rsts from control measurements. To
identify packet drop based censorship we look for the absence of a payload
response in addition to the presence of a payload response for the controls.
Differentiating true censorship from network phenomena for both these forms of
censorship can be challenging as: 1) \rst packets can occur for reasons
unrelated to censorship, 2) lack of \rst packets or payload response can occur
due to packet loss, and 3) transient censorship failures can cause \rst packets
to be missing or payload responses to appear. 

To disambiguate censorship from network behavior we perform repeated
measurements separated by a fixed time interval. We note a significant time
interval needed to ensure the tuple can be reused by the remote operating
system. We use 30 minutes, which is 15x longer than the RFC recommendation~\cite{rfc793}
and significantly longer than needed based on our experiments. We note
long-lived residual censorship~\cite{bock2021your} beyond 30-minutes is not a concern,
since our goal is to study routes, not censor activity.
We only consider a result censorship for a particular (destination,
source) parameter if: 1) we obtained a \rst for all the repetitions and
\emph{no} \rst for any of the control measurements, or 2) we obtained \emph{no}
payload responses for any of the repetition and responses for all the
control measurements. We only consider a result \emph{not} censorship if across all
repetitions: 1) no \rst packet was observed for \rst based censorship, or 2)
payload responses were obtained for packet drop based censorship. We exclude
all other results. This approach is \emph{conservative}, as it gives us a lower
bound of possible true-positive results, discarding potential differences
due to network changes, packet loss, etc.  
\ie this approach gives us a set of differences we have high confidence
are routing-induced, rather than transient effects. Other phenomena and instances
of routing-induced censorship differences may very well exist and be excluded by
this method due to transient network effects, but our results will still
provide a lower-bound.

Blockpages are detected by conducting a manual study of
censorship in each country to extract blockpage templates
that are then matched against responses. This
manual operation ensures matches with no false positives but possible
false negatives. This is acceptable, as it provides a lower bound on
our findings on routing-induced censorship changes, rather than
exhaustive censorship measurement. 
The lack of a blockpage could also be attributed to similar
causes as the lack of a \rst packet. To account for this we perform
similar repetitions and consider a result valid when there is
consistency across all iterations.

\PP{UDP-based censorship.} 
For \udp censorship, \name extends prior work~\cite{breadcrumb}.
We send route-stable sensitive and control \dns
\texttt{A?} queries to destinations globally that \emph{do not} operate DNS
resolvers and record all responses.  Censorship occurs when we obtain a
response to a sensitive query, but not the control.

\PP{Improving Result Reliability.}
Across all experiments we use a combination of control measurements
and repetition separated by fixed time intervals to obtain
reliability.  We only consider censorship present or absent for a
particular experiment if the relevant behavior is present
\textbf{across all} repetitions in addition to consideration of
control experiments. This stabilizes our results by eliminating
factors like packet loss, residual censorship, and transient behaviors
that may appear as censorship changes. It is possible that within
the time-frame of our repetitions network topologies may change,
resulting in a missed non-censorship result. This is acceptable as our
goal is to have a lower bound on censorship changes.

\PP{Route-Stable, Per-Protocol Traceroute.} Beyond simple observed
censorship differences, we also seek to understand \emph{why} these
differences occur, which requires constructing the routes taken by
\emph{specific} packets with \emph{specific} parameters.  Common prior
traceroute techniques are built on \udp, \tcp-based methods use
\syn packets, and purpose built censorship traceroute
techniques~\cite{raman2022network} do not control for \flowid.
For our usage we need to measure the path of
\emph{specific} sensitive packets with full control over all aspects
of the packets; no tools allow reconstructing routes of a specific
user-controlled packet. Further, for \tcp, the sensitive packet is
sent after establishing a \tcp connection, thus we must iterate the \ttl of the
payload packet of an active connection. This is challenging as we cannot
control retransmission or \tcp state from user space. 
Thus we develop a
route-stable traceroute tool to measure the path taken by packets for
all protocols we explore. We control all fields in the packet that
influence routing, varying the \ttl of a given sensitive packet, and
build the path using \texttt{ICMP} responses.

\label{subsec:traceroute}

In lieu of implementing a \tcp stack in user-space to vary \ttls
of mid-flow sensitive packets, \name leverages a
combination of Netfilter nfqueues~\cite{nfqueue} and firewall rules.
Our firewall redirects outbound packets we seek to traceroute
to an nfqueue, and our nfqueue hook sends copies of the packet at incrementing
\ttls. Kernel retransmission attempts are dropped until the traceroute is
complete or the connection is terminated by the endpoint or censor. 
This technique ensures: 1)
we can capture the actual path taken by the packets, and 2) it can be
applied to any censorship measurement technique. With this technique the
connection does not need to be torn down after each \ttl attempt, as the packet
does not reach the end-host until the end of the trace. We embed the \ttl in the \texttt{IP ID} field of the
packet to disambiguate responses. We note the \texttt{IP ID} field is not known to be used for
\flowid~\cite{paris_traceroute}, and does not influence route.
We extract the source parameters and \ttl from the packet embedded in the
\texttt{ICMP} response to build the network path.

\subsection{Ethical Considerations}
\label{sec:ethics}
Measuring Internet censorship requires careful consideration of the ethical
implications of all experiments, and weighing those implications against the
potential benefits of the understanding gained from the experiment. We build
our ethical framework around the models from prior censorship measurement
works~\cite{breadcrumb,augur,iris}, who in turn modeled their work after the
Belmont~\cite{belmont} and Menlo~\cite{menlo} reports. Namely, we consider the
concepts of \emph{justice}, \emph{respect for persons}, \emph{beneficence}, and
\emph{respect for law and public interest}. Broadly speaking, these principles
dictate that censorship measurement researchers should strive to: 1) Ensure
those who bear the risk of the work are also the work's beneficiaries, 2) Given
the impossibility of obtaining informed consent, seek to \emph{minimize} risk,
3) Ensure that no experiments stress the infrastructure or users' machines.

We call attention to \emph{beneficence} which deals with experimentation that has inherent risks and speaks to the need to reduce
risk to the extent such that the benefits of conducting the measurements
outweigh the risk. Prior censorship measurement~\cite{iris,augur,breadcrumb} discussed this concept. \eg ``In lieu of
attempting to obtain informed consent, we turn to the principle of beneficence,
which weighs the benefits of conducting an experiment against the risks
associated with the experiment. Note that the goal of beneficence is not to
eliminate risk, but merely to reduce it to the extent possible.''~\cite{iris}.

Guided by these principles, we reduce risk by: 1)
significantly down sampling measurement vantage points to a
minimum set per autonomous system necessary to show the effect of \ecmp on
censorship measurement, 2) measuring only a single censored domain per host
(the same domain across an entire country) which we manually selected to
minimize potential harm (\eg by excluding terrorism or similar categories), 3)
only conducting remote measurements that do not result in follow-on
host-initiated communication with censored domains or IPs, and 4) we rate-limit
and randomize experiments to minimize load on any machine. 

We note that the benefits of this work include providing the community with the
knowledge of how to conduct sound censorship measurement, which will enable us
to develop tools that better understand censorship globally, with fewer
measurements.  Such understanding in-turn enables the development of better
circumvention technologies, and aids policy makers and activists; all benefits
which impact the population that bears the risk of the experiments.

\section{Research Questions and Experiments}
\label{sec:dataset}
\label{dataset}

We begin by enumerating our research questions, and then defining the experiments we designed
and datasets we collected to answer them.
We seek to answer these questions across protocols, censorship techniques,
and countries:
\begin{itemize}[leftmargin=1em, nosep]
    \item \textbf{RQ1:} \rqone
    \item \textbf{RQ2:} \rqtwo
    \item \textbf{RQ3:} \rqthree
    \item \textbf{RQ4:} \rqfour
    \item \textbf{RQ5:} \rqfive
\end{itemize}

\noindent These research questions build on one another, beginning
with an exploration of the impact of varying source parameters (and thus \flowid) on the path measurement packets
take to vantages globally,
regardless of censorship, and then observing this impact on censorship
results. We then perform a deeper analysis to understand the
different reasons that cause such variation. We end with trying to
understand the applicability of such variation on prior censorship
measurement studies, and its potential impact on two specific studies. All experiments were
conducted from a purpose built /24 scanning network at an academic
institution in North America, using the methods from Section~\ref{sec:Method}.

\subsection{RQ1: \rqone}
\label{sec:exp1}
In \textbf{RQ1} we explore how varying censorship measurement packet source IP
and source port changes paths across countries and protocols, absent censorship. The
goal in exploring this question is quantifying the extent to which different
parameters result in different routes, which is a prerequisite for \ecmp
censorship measurement differences.
To answer this, we conduct route-stable
\http, \https, and \dns traceroute measurements
(Section~\ref{subsec:traceroute}) with a control domain to a geographically
diverse set of vantage points in various countries. Across measurements
we either fix source parameters, vary an individual parameter, or vary both
parameters~\cite{breadcrumb}.%

\PP{Destination IPs.} To understand the impact of
packet parameters on route, we select a
diverse set of measurement IPs within each country. Starting with a
list of all possible destinations for each protocol, we use Censys~\cite{censys} data
to extract destinations with
port 80 open for \http, port 443 open for \https and
port 53 \emph{not} open for \dns). We then pick one destination per AS for each country
tested. The number of destinations selected will
vary for each country.%

\PP{Domain.} As our goal in this scenario is to measure the
variation in packet path in different censorship measurement scenarios
and not to perform censorship measurement, we perform measurements
with a known
benign domain (\nolinkurl{example.com}). The advantage of this is
twofold: (1) keeping with our ethical considerations, we reduce risk to
individuals by using benign domains when possible, and (2) we avoid
any side-effects from censorship activity that could possibly alter
the state of open connections used for traceroute (\eg \rst packets that tear down active
connections).

\PP{Parameter Variation} When selecting source IPs and ports,
we conduct
four variations of the following experiments:

\begin{itemize}[leftmargin=1em,nosep]
\item \textbf{Everything Constant. } We repeat the experiment 144
  times, replicating experiments from censorship measurement
  tools across \dns, \http, and \https~\cite{quack, iris, bock2021even}. 
  
\item \textbf{Varying Source Port. } Varying the source port of the
  measurement packet with 144 randomly selected ports from the
  ephemeral range and fixing the source IP.
  
\item \textbf{Varying Source IP. } Varying the source IP of the
  measurement packet with 144 randomly selected IPs from a \emph{/24}
  subnet and using a fixed source port.

\item \textbf{Varying Source IP and Source Port. } Varying both the
  source IP and source port of the packet simultaneously. We pick 12
  randomly selected source IPs and source ports, ensuring that we
  obtain 144 measurements in total.
\end{itemize}

\noindent This four-fold approach gives us the ability to understand the variation of the
path taken by the packet across different individual dimensions and with them
combined. Of note is the need to ensure a consistent total number of
experiments (\eg 144 experiments vs 12 experiments 12 times), to ensure
comparisons \emph{between} experiments are apt. We pick 12 based on
prior work~\cite{breadcrumb} which found that results stabilized by
iteration 12 (Section 4.5~\cite{breadcrumb}).

\subsection{RQ2: \rqtwo}
\label{sec:exp2}
Next \textbf{RQ2} seeks to explore the impact of \ecmp routing on the
\emph{results} of outside-in censorship measurement across protocols, censorship
mechanisms, and countries.  \ie do different routes lead to different
censorship results? We thus conduct an experiment utilizing the methods
described in Section~\ref{sec:Method}, varying different source parameters,
across numerous censorship methods and countries.  We do not perform
traceroutes for this experiment, reducing packets
and risk.  %

\PP{Country and Domain Selection.}
For this experiment our goal is to find all potential countries with
state-sponsored and ISP censorship mechanisms amenable to outside-in
measurement, and an associated sensitive domain that can be used to
measure the same. We do this by performing a preliminary manual
exploration on \emph{all} countries and dependant territories we could identify as having
IP addresses geolocated too. This totaled 249 countries and dependent territories.
For each, we: 1) pick potential
sensitive domains using a combination of OONI~\cite{ooni} and
CLBL~\cite{clbl}, and sample destination IPs (open on 80 and 443, and
closed on 53, similar to RQ1) from Censys~\cite{censys}, and 2) perform a handful of external
censorship measurements to these destination IPs on the
different potential sensitive domains.
As a result of this
exploration, we find 21 countries from Europe, Asia, Africa, and
Middle East that were able to be remotely measured. We explore this RQ
with the associated domains that are confirmed to be censored in the
target country.
We note that such an approach may not exhaustively find
\emph{all} countries and protocols that are amenable to outside-in
measurement, but this is acceptable, as our goal is to understand and provide a lower bound on
the
phenomenon's generalizability globally, not exhaustively enumerate all possible scenarios.

\PP{Destination IPs}
\label{subsec:destipselection}
For this experiment, for each country, we select a set of geographically
diverse destination IPs to get an accurate picture of the change in censorship
due to \ecmp routing. The criteria we look for in a
destination IP are unchanged from RQ1: open on 80 and 443, and closed on 53.
These criteria allow
us to perform measurements across protocols \http, \https, and
\dns on the same IP.
From this usable set of IPs (identified from
Censys~\cite{censys}), we sample up to 60 destinations per AS,
depending on the volume of ASes and available IPs for each country. AS
mapping is performed with the RouteViews prefix-to-as dataset~\cite{pfx2as}.

\PP{Source Parameters.} We use 208
source IPs selected from a single \texttt{/24} subnet (excluding
\texttt{.0} and \texttt{.255}), and 8 source
ports randomly selected from the ephemeral port range. We ensure
that the source IPs are normally distributed with respect to their lowest 3
bits, which is relevant for quantifying some load balancing~\cite{cisco_load_balance}.

\subsection{RQ3: \rqthree}

\textbf{RQ3} seeks to understand whether particular source parameters (IP
or port) have a noticeable impact on variation in censorship results caused by
\ecmp routing, \ie do some source IPs/ports have a greater impact than others?
Exploring this begins to sheds light \emph{why}, from a packet perspective, differences occur, and potentially
identifies different \ecmp algorithms that use different parts of the packet to
perform routing.  We use measurements from RQ2's experiments
to answer this question and explore impact across source IPs, source ports and
their combinations.

\subsection{RQ4: \rqfour}

Next \textbf{RQ4} explores different underlying causes
for \ecmp routing induced censorship differences. \ie is \ecmp routing
causing sensitive packets to route around censorship, traverse
through failed censoring nodes, or pass through completely different
geographical locations with different censorship behavior? To answer
this question we conduct route-stable traceroute measurements
(Section~\ref{subsec:traceroute}) to produce network graphs which we
use to answer this question. We describe the dataset used
subsequently:

\PP{Destinations and Sources}
We use the measurements of \textbf{RQ2} to sample destinations
for all countries/protocols that exhibit variation and produce network
graphs for those source parameters (Section~\ref{subsec:traceroute}). 
We select destination and source
parameter combinations that exhibited consistent,
repeatable censorship change. 

Our goal is to: 1) qualitatively demonstrate that the differing
observed censorship results are a direct consequence of varying
network paths, and 2) understand and quantify the different underlying
effects that contribute to such variation. We perform this exploration
across all censorship methods and countries from RQ2.

\subsection{RQ5: \rqfive}
The goal of \textbf{RQ5} is twofold: 1) to understand the
applicability of \ecmp routing on the variation observed in prior studies,
and 2) to understand the potential impact \ecmp routing had on 2 specific prior global outside-in
censorship measurement studies. For the later, we focus on censorship differences
unattributed in prior work that could be \ecmp routing induced. We
stress that these 2 prior works are not \emph{incorrect}, but rather effects
observed in such work may be attributable to \ecmp routing, rather
than other effects. We seek to understand the impact of \ecmp routing
induced censorship differences in two dimensions: 1) can unattributed
ambiguity in prior results be potentially attributed to \ecmp routing,
and 2) are end-to-end results from prior work impacted by \ecmp
routing? We use the measurements of \textbf{RQ2} to answer this
questions.

When comparing to the 2 prior studies, we make a best-effort to use their
published measurement methods to sample the same amount of
destinations from the same locations, and compare those results with
our own.  We do not replicate their \emph{specific} destinations, due
to both the significant time since their studies were collected during
which routes likely changed, and the dataset of specific IPs not being
publicly available.  Our formulation is meant to understand potential
impact and scope, rather than identify precise results.

\section{Results}
\label{sec:results}

\begin{table}[]
 \resizebox{\columnwidth}{!}{%
  \begin{tabular}{|l| cc | cc | cc |}
    \hline
    & \multicolumn{2}{c|}{\dns}                    & \multicolumn{2}{c|}{\http}                   & \multicolumn{2}{c|}{\https}                                        \\
    \hline
    \multirow{2}{*}{Country} & \multicolumn{2}{c|}{\multirowcell{2}{Methods \&\\ Affected?}} & \multicolumn{2}{c|}{\multirowcell{2}{Methods \&\\ Affected?}} & \multicolumn{2}{c|}{\multirowcell{2}{Methods \&\\ Affected?}}\\
    & & & & & & \\ \hline
    Algeria (DZ)      &  \multicolumn{1}{c|}{-}  & -        &  \multicolumn{1}{c|}{Drop+\rstnospace}   & \cmark    &  \multicolumn{1}{c|}{Drop}   & \cmark     \\
    Bangla. (BD)     &  \multicolumn{1}{c|}{-}     & -         &  \multicolumn{1}{c|}{BPage+Drop}   & \cmark    &  \multicolumn{1}{c|}{Drop+\rstnospace}   & \cmark     \\
    Belarus (BY)     &   \multicolumn{1}{c|}{Inject.}   & \xmark    &  \multicolumn{1}{c|}{BPage}   & \cmark    &  \multicolumn{1}{c|}{\rstnospace}   & \cmark     \\
    China (CN)    &  \multicolumn{1}{c|}{Inject.}     & \cmark         & \multicolumn{1}{c|}{\rstnospace}     & \cmark         & \multicolumn{1}{c|}{\rstnospace}     & \cmark          \\
    India (IN)     &   \multicolumn{1}{c|}{Inject.}   & \xmark    &  \multicolumn{1}{c|}{BPage+\rstnospace}   & \cmark    &  \multicolumn{1}{c|}{\rstnospace}   & \cmark     \\
    Indonesia (ID)     &  \multicolumn{1}{c|}{-}     & -         &  \multicolumn{1}{c|}{BPage}   & \cmark    &  \multicolumn{1}{c|}{\rstnospace}   & \cmark     \\
    Iran (IR)      &   \multicolumn{1}{c|}{Inject.}   & \cmark    &  \multicolumn{1}{c|}{BPage+\rstnospace}   & \cmark    &  \multicolumn{1}{c|}{Drop+\rstnospace}   & \cmark     \\
    Jordan (JO)     &  \multicolumn{1}{c|}{-}     & -         &  \multicolumn{1}{c|}{\rstnospace}   & \xmark    &  \multicolumn{1}{c|}{\rstnospace}   & \cmark     \\
    Kuwait (KW)     &  \multicolumn{1}{c|}{-}     & -         &  \multicolumn{1}{c|}{Drop+\rstnospace}   & \cmark    &  \multicolumn{1}{c|}{Drop+\rstnospace}   & \cmark     \\
    Oman (OM)    &   \multicolumn{1}{c|}{Inject.}   & \xmark    &  \multicolumn{1}{c|}{BPage+Drop}   & \xmark    &  \multicolumn{1}{c|}{\rstnospace}   & \cmark     \\
    Pakistan (PK)    &   \multicolumn{1}{c|}{Inject.}   & \xmark    &  \multicolumn{1}{c|}{Drop+\rstnospace}   & \cmark    &  \multicolumn{1}{c|}{Drop+\rstnospace}   & \cmark     \\
    Qatar (QA)      &  \multicolumn{1}{c|}{-}     & -         &  \multicolumn{1}{c|}{BPage}   & \xmark    &  \multicolumn{1}{c|}{\rstnospace}   & \xmark     \\
    Russia (RU)     &   \multicolumn{1}{c|}{Inject.}   & \cmark    &  \multicolumn{1}{c|}{BPage+Drop}   & \cmark    &  \multicolumn{1}{c|}{Drop+\rstnospace}   & \cmark     \\
    Rwanda (RW)     &  \multicolumn{1}{c|}{-}     & -         &  \multicolumn{1}{c|}{\rstnospace}   & \cmark    &  \multicolumn{1}{c|}{\rstnospace}   & \cmark     \\
    S. Korea (KR)     &  \multicolumn{1}{c|}{-}     & -         &  \multicolumn{1}{c|}{BPage}   & \cmark    &  \multicolumn{1}{c|}{\rstnospace}   & \cmark     \\
    Syria (SY)      &  \multicolumn{1}{c|}{-}     & -         &  \multicolumn{1}{c|}{\rstnospace}   & \cmark    &  \multicolumn{1}{c|}{\rstnospace}   & \cmark     \\
    Turkey (TR)     &   \multicolumn{1}{c|}{Inject.}   & \xmark    &  \multicolumn{1}{c|}{\rstnospace}   & \cmark    &  \multicolumn{1}{c|}{\rstnospace}   & \cmark     \\
    Turkmen. (TM)      &   \multicolumn{1}{c|}{Inject.}   & \cmark    &  \multicolumn{1}{c|}{\rstnospace}   & \xmark    & \multicolumn{1}{c|}{-}     & -          \\
    UAE (AE)    &  \multicolumn{1}{c|}{-}     & -         &  \multicolumn{1}{c|}{BPage}   & \xmark    & \multicolumn{1}{c|}{-}     & -          \\
    Uzbek. (UZ)      &  \multicolumn{1}{c|}{-}     & -         &  \multicolumn{1}{c|}{BPage}   & \xmark    &  \multicolumn{1}{c|}{Drop+\rstnospace}   & \xmark     \\
    Yemen (YE)      &  \multicolumn{1}{c|}{-}     & -         &  \multicolumn{1}{c|}{\rstnospace}   & \xmark    &  \multicolumn{1}{c|}{\rstnospace}   & \xmark     \\
    \hline
  \end{tabular}
  }
  \caption{
    Impact of \ecmp routing on censorship measurement.
    17 out of the 21 countries show changes in
    due to route. \cmark~denotes changes,
    \xmark~denotes no changes, and - denotes no such externally
    measurable censorship.
  }
  \label{tab:resultsoverview}
  \vspace{-0.75em}
\end{table}

We now explore results from the experiments outlined in
Section~\ref{sec:dataset}, focused on answering the \rqnum motivating research
questions.
Table~\ref{tab:resultsoverview} provides an overview of results from our study
detailing the censorship explored for each country.  We
find 17 out of 21 countries explored that perform externally measurable state-level censorship
are subject to variation in observed
censorship results due to \ecmp.

\subsection{RQ1: \rqone}
\label{sec:rq1}

\begin{figure}[]
  \centering
  \includegraphics[width=\columnwidth]{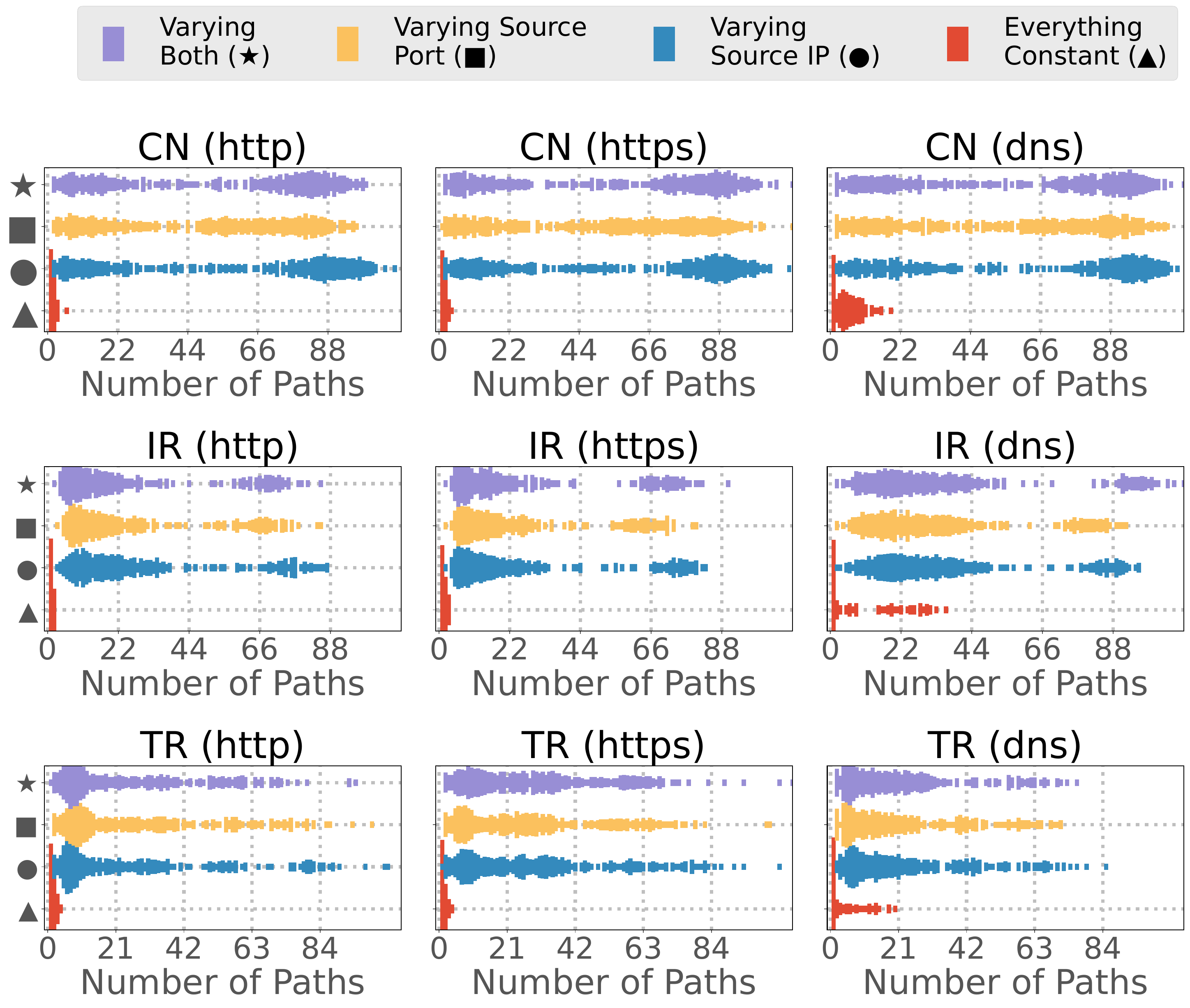}
    \caption{\emph{Normalized Distribution of Number Of Paths for all
      destinations.} Marker size represents number of experiments
      across all destinations that had the particular \emph{Number of Paths}.
      We observe that the variation in path has two observable
    modes, constant parameters vs varied parameters, and that source IP has 
    a greater impact on network path variations than varying source port.  We
    also observe different modes within a particular country (\eg Iran)
    potentially indicating destination port-specific based load balancing.}
  \label{fig:paths}
\end{figure}

We aim to
understand if varying networking paths leads to changes in censorship measurement globally, across protocols and ports.  We employ the
metric of \textit{Number of Nodes} and \textit{Number of Paths} to capture
variation in network path.  \textit{Number of Nodes}, for a destination is a
\emph{set} of all unique router hops in the network path across all
experiments. A \emph{path} is defined to be a set of all nodes for a particular
combination of destination and source parameters.  \textit{Number of Paths} is
thus a count of the set of all paths for a destination.

This form of \ecmp routing within a country potentially depends on the
network infrastructure of the country and its ISPs.  Additionally, as
packets pass through several different infrastructures to reach a destination,
they pass through several different ASes~\cite{dminer} with potentially different network
infrastructures that can influence their path. Similarly routing may differ
based on destination port~\cite{paris_traceroute}, causing different protocols
to result in different routing. 

Figure~\ref{fig:paths} shows results from these experiments
(Section~\ref{sec:exp1}) and the variation in network path as a function of
varying different source parameters part of the \flowid.  Results are
represented as the (normalized) distribution of the \emph{Number of Paths} for
the experiments conducted for each country and protocol.
For space we
abbreviate the number of countries shown to a characteristic subset.
Appendix~\ref{sec:appendix_variation} shows
similar results for \textit{Number of Nodes}.

For the experiment where we fix source parameters and repeat the
measurement (144 times), we see that there is very little to no
variation in the network path. The mean \emph{Number of Paths} remained $\leq$ 3
across all the different protocols tested in different
countries. This observation aligns with expectations and prior work~\cite{breadcrumb}
and holds for a larger set of protocols and countries. Although the
path remains constant across repetitions most of the time, there can
be subtle variation over time in some cases (\eg TR and IR with \dns).
We find that repeating the experiments with the same source
parameters does not frequently exercise different network paths.

Figure~\ref{fig:paths} shows that varying the different source parameters
(independently and together) yields notably different behavior in the network
path variation among different countries. For a (country, protocol), we observe
two general modes in the observed variation: 1) varying source IP has a
greater effect on the network path than varying source port (\eg CN \& IR), 
and 2) varying source IP/port seems to have a similar effect on
resulting path density (\eg TR). 
In addition, we see modes in the
observed variation pointing to different infrastructure
(\eg IR, TR, CN), highlighting the need for both source IP
\textit{and} port variation.

Apart from variation between countries, we also observe variation in path
diversity across \emph{different protocols} within the same country. In the
case of Iran, we see that the distribution of \emph{Number of Nodes} varies
between \http, \https \& \dns. We hypothesize this steams from destination port
also being used for \ecmp routing, opening the door to differences in
censorship measurement based solely on port. All told, we note the significance
in path diversity in censored countries due to source parameters indicates
censorship variations are possible, explored next.

\subsection{RQ2: \rqtwo}
\label{sec:rq2}

RQ1 established that varying fields contributing to \flowid impacts the network
path taken by a censorship measurement packet for numerous protocols across various
countries.  We now look at \ecmp routing's impact on observed
censorship results globally. 
For this experiment we selected a small number of destinations per AS per
country (Section~\ref{subsec:destipselection}).  Per our ethical
framework (Section~\ref{sec:ethics}), our goal is to limit our measurements while still
confirming the existence of the phenomenon.  We then conduct measurements across
different protocols with the same set of destinations and source parameters,
for one control domain and one sensitive domain specifically picked to exercise
censorship in the country.  We use the methods described in
Section~\ref{sec:method} to identify the presence or absence of \rsts,
intentional packet drops, blockpage, or \dns based censorship.  We then
identify destinations that produce different observed censorship for different sets of source parameters.

\begin{table}[]
  \centering
    \begin{tabular}{|c|l|l|l|}
      \hline
      \textbf{Country} & \textbf{Protocol} & \bftwolinecell{Destinations\\(ASes)} & \bftwolinecell{Destinations\\(ASes) Affected}\\
      \hline
      \multirowcell{2}{Algeria}    &   \http   &    311 (4)   &    2\% (25\%)\\
      &   \https   &    288 (5)   &    3\% (40\%)\\
      \hline
      \multirowcell{2}{Bangladesh}    &   \http   &    743 (119)   &    41\% (58\%)\\
      &   \https   &    750 (122)   &    37\% (54\%)\\
      \hline
      \multirowcell{2}{Belarus}    &   \http   &    1094 (13)   &    12\% (31\%)\\
      &   \https   &    1049 (10)   &    11\% (20\%)\\
      \hline
      &   \dns   &    2501 (289)   &    25\% (46\%)\\
      China    &   \http~\tablefootnote{\label{fn:china}                                                                                               
        During our study a significant change to                                                                         
        China's GFW was deployed, resulting in a new form of reactive                                                                         
        blocking~\cite{china2022blocking}, impacting our ability to measure \ecmp routing in China.  Thus we present partial results for HTTP and HTTPS                                                                   
        predating the change.}   &    210 (64)   &    23\% (21\%)\\
      &   \https~\textsuperscript{\labelcref{fn:china}}   &    948 (217)   &    69\% (71\%)\\
      \hline
      \multirowcell{2}{India}    &   \http   &    3289 (87)   &    99\% (99\%)\\
      &   \https   &    3316 (90)   &    100\% (100\%)\\
      \hline
      \multirowcell{2}{Indonesia}    &   \http   &    974 (78)   &    84\% (71\%)\\
      &   \https   &    979 (80)   &    84\% (68\%)\\
      \hline
      &   \dns   &    1774 (127)   &    9\% (26\%)\\
      Iran    &   \http   &    1351 (142)   &    2\% (2\%)\\
      &   \https   &    1347 (139)   &    9\% (20\%)\\
      \hline
      Jordan    &   \https   &    231 (4)   &    $<$ 1\% (25\%)\\
      \hline
      \multirowcell{2}{Kuwait}    &   \http   &    1188 (15)   &    91\% (80\%)\\
      &   \https   &    70 (11)   &    90\% (82\%)\\
      \hline
      Oman    &   \https   &    842 (15)   &    2\% (7\%)\\
      \hline
      \multirowcell{2}{Pakistan}    &   \http   &    373 (52)   &    15\% (18\%)\\
      &   \https   &    704 (93)   &    44\% (62\%)\\
      \hline
       \multirowcell{3}{Russia} &   \dns   &    105 (33)   &    65\% (46\%)\\
         &   \http   &    1808 (327)   &    66\% (47\%)\\
      &   \https   &    2526 (362)   &    15\% (20\%)\\
      \hline
      Rwanda    &   \http   &    23 (1)   &    92\% (100\%)\\
      &   \https   &    54 (1)   &    75\% (100\%)\\
      \hline
      \multirowcell{2}{South Korea}    &   \http   &    951 (44)   &    17\% (44\%)\\
      &   \https   &    391 (29)   &    2\% (4\%)\\
      \hline
      \multirowcell{2}{Syria}    &   \http   &    154 (1)   &    14\% (100\%)\\
      &   \https   &    152 (1)   &    15\% (100\%)\\
      \hline
      \multirowcell{2}{Turkey}    &   \http   &    526 (36)   &    21\% (34\%)\\
      &   \https   &    410 (33)   &    18\% (31\%)\\
      \hline
      Turkmenistan    &   \dns   &    643 (1)   &    5\% (100\%)\\
      \hline
    \end{tabular}
\caption{RQ2 Summary. Given are number of censored destinations studied
 (number of ASes in parentheses) and
  percent impacted by \ecmp; impact varies significantly by context.
}
\label{tab:rq2}
\end{table}

\begin{figure}[]
  \centering
  \includegraphics[width=\columnwidth]{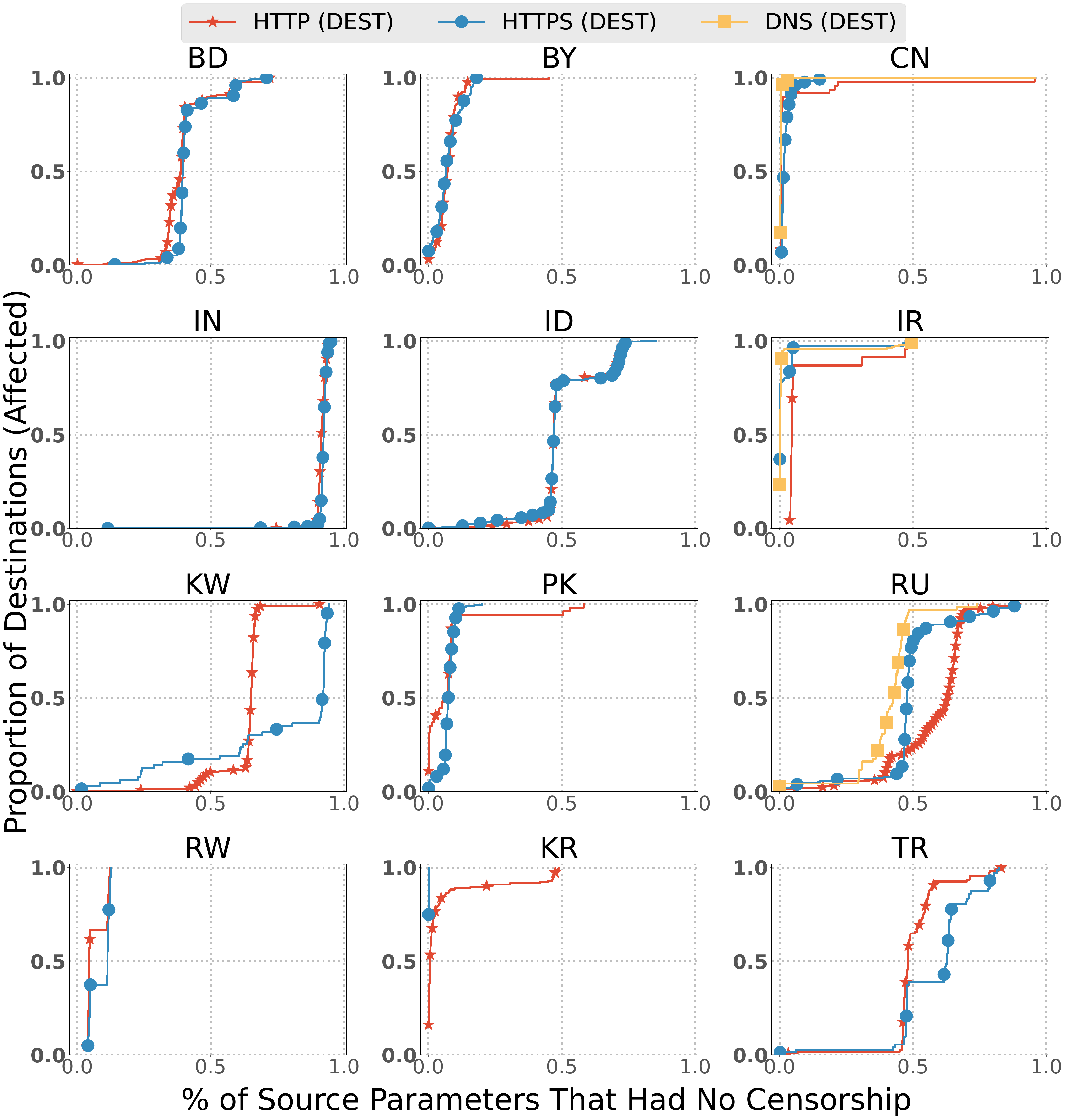}
  \caption{CDF of percent of (source IP, source Port) combinations per destination for
    which we observed \textbf{no} censorship. Measurements are limited to the subset of destinations
    where variation is observed in Table~\ref{tab:rq2}.
    We observe that 1) the percent of source
    parameters that produce no censorship for the median affected
    destination varies significant by country and protocol. 
    2) \http and \https follow very similar trends in how source
    parameters affect their results in some countries (BD, BY, IN, \& ID)
    while in others, they vary significantly (RU, KW)
    and 3) \dns is affected least by source parameters.
    }
  \label{fig:oddifacepcent}
\end{figure}

\textbf{We find that censorship measurement from 17 out of 21 countries are
affected by \ecmp routing, with 34 out of 49 contexts impacted.}
Table~\ref{tab:resultsoverview} provides results across all
countries and protocols.  
We find that the extent of impact varies
significantly with country \textit{and} protocol, ranging from
\changesRangeHighHTTP to \changesRangeLowHTTP of destinations (depending on
country) affected for \http, from \changesRangeHighHTTPS to
\changesRangeLowHTTPS of destinations affected for \https, and from
\changesRangeHighDNS to \changesRangeLowDNS of destinations affected for \dns.
Table~\ref{tab:rq2} provides a breakdown of these results. Routing
impact not only varies for different countries, but also for different
protocols within a country. \eg \exampleCountry shows differences across
\examplehttpChange, \examplehttpsChange, and \examplednsChange of
destinations for \http, \https, and \dns respectively.
Such comparison among protocols supports the
hypothesis of different underlying censorship infrastructures possibly deployed at
different locations in addition to
\ecmp routing affecting protocols differently.  
We speculate our observed lower
Chinese \dns censorship impact compared to prior
work~\cite{breadcrumb} is due to our
use of fewer parameters.
Appendix~\ref{sec:appendix_overlap} 
shows overlap across protocols.

\subsubsection{Prevalence of Source Parameters Yielding Changes}
We now explore measurements for each of
the censorship protocols across the different countries, by number of destinations and source parameters. We seek to understand
the prevalence of parameters leading to changes in censorship 
responses in order to gauge the expected impact of \ecmp routing on censorship
measurement. \eg do many, or only a few parameters, impact results?

\begin{figure*}[]
  \centering
  \subfloat[China (\https)\label{fig:china-https}]{\includegraphics[width=0.3\textwidth]{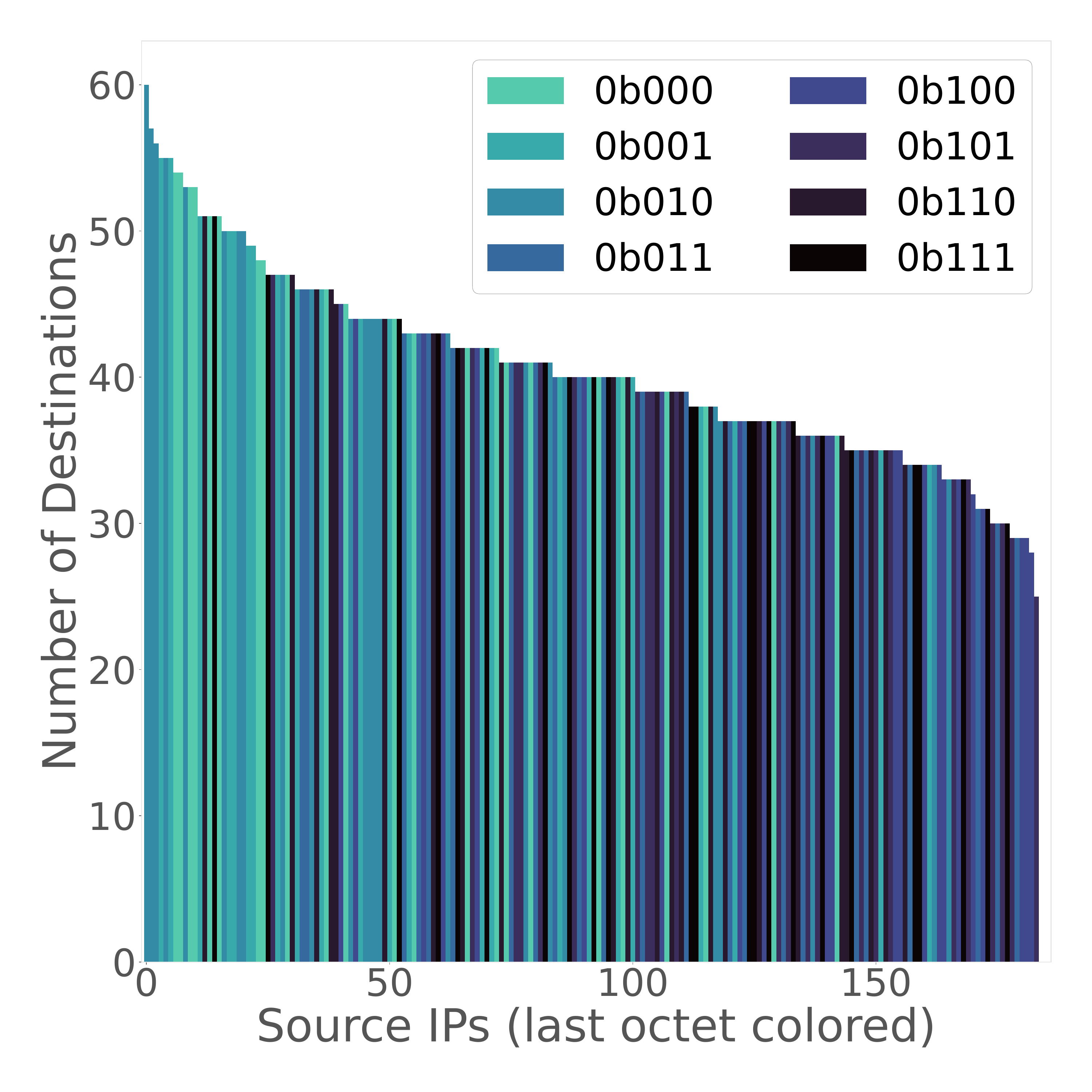}}\hfill
  \subfloat[Syria (\http)\label{fig:syria-http}]{\includegraphics[width=0.3\textwidth]{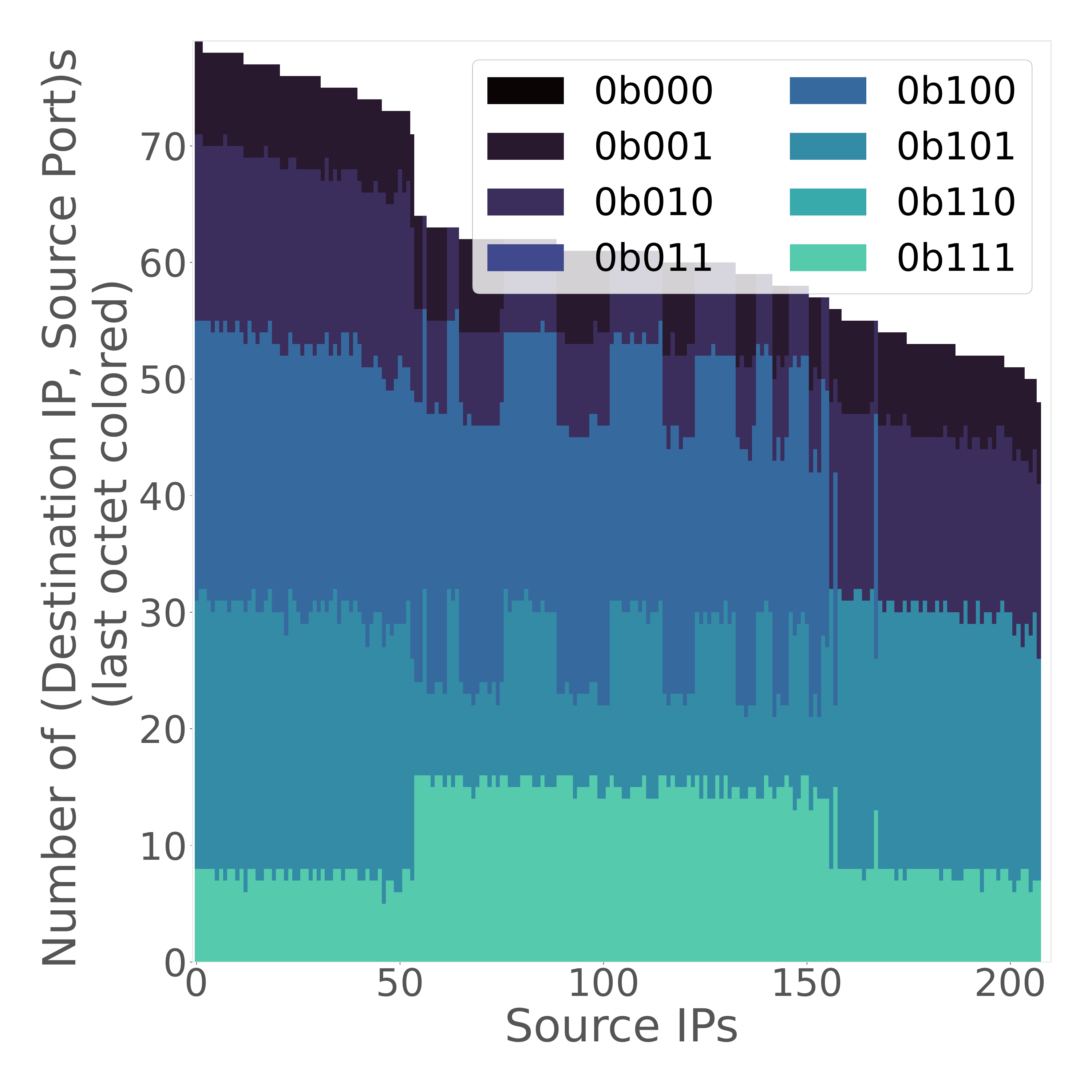}}\hfill
  \subfloat[South Korea (\http)\label{fig:kr-http}]{\includegraphics[width=0.3\textwidth]{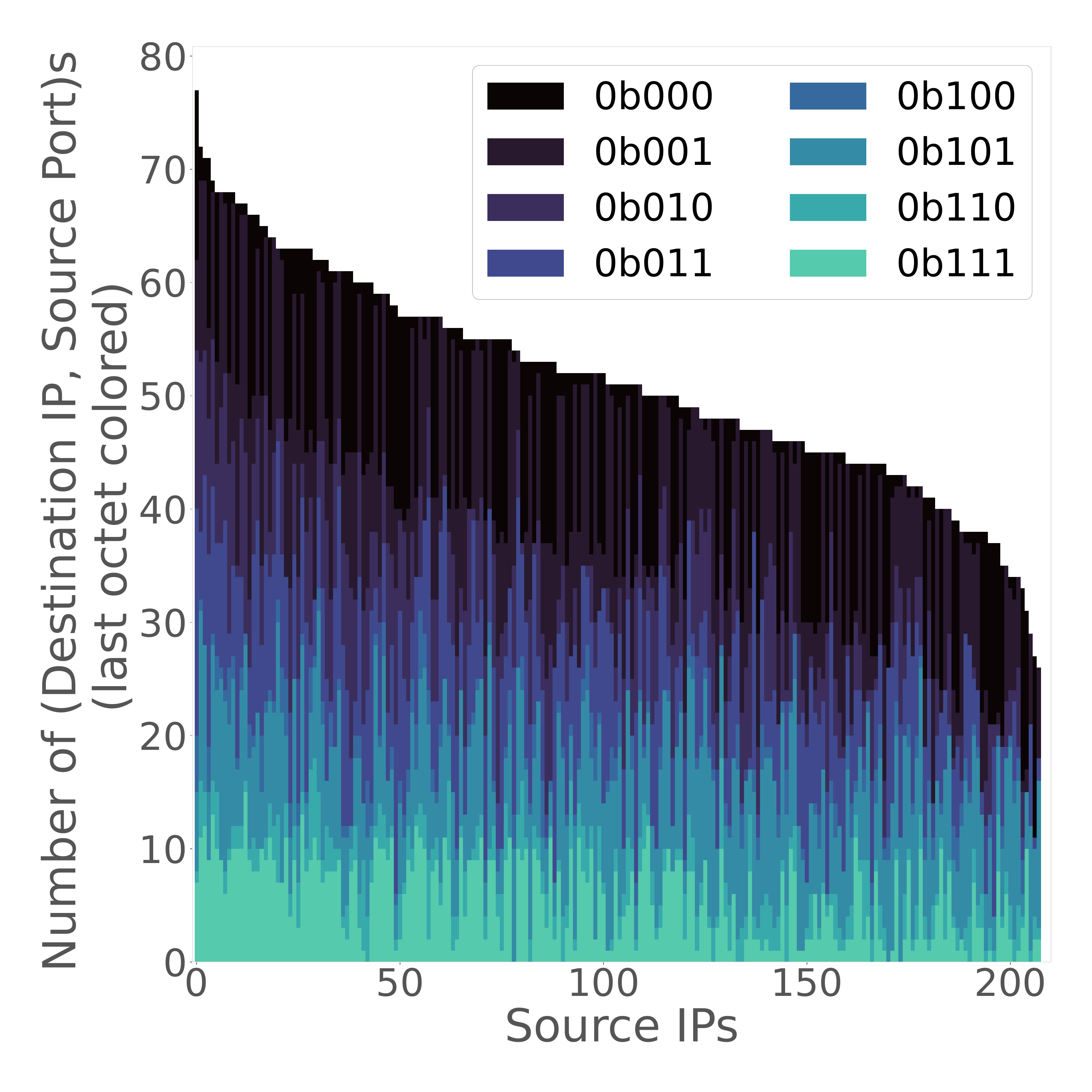}}

  \subfloat[India (\http)\label{fig:in-http}]{\includegraphics[width=0.3\textwidth]{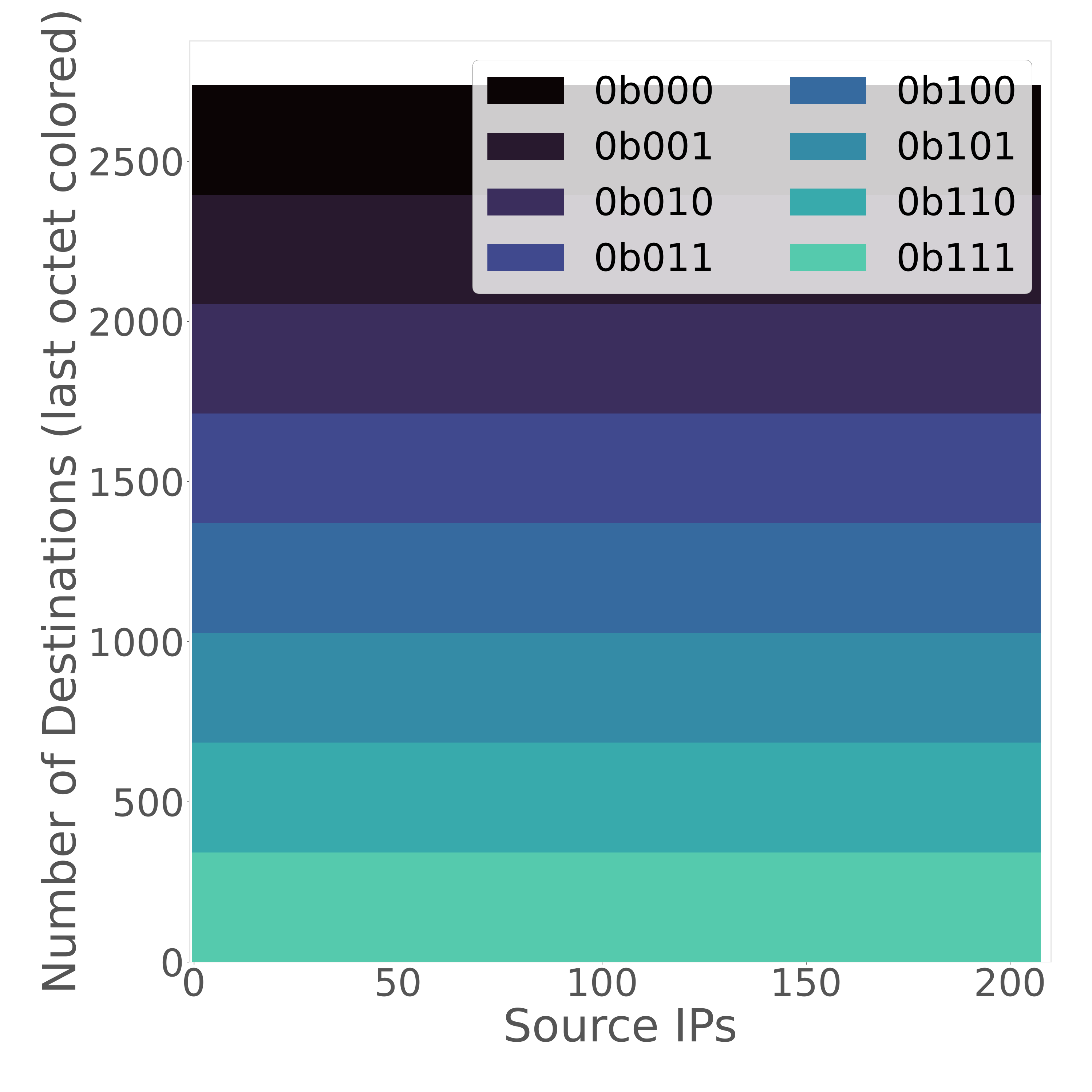}}\hfill
  \subfloat[Bangladesh (\http)\label{fig:bd-http}]{\includegraphics[width=0.3\textwidth]{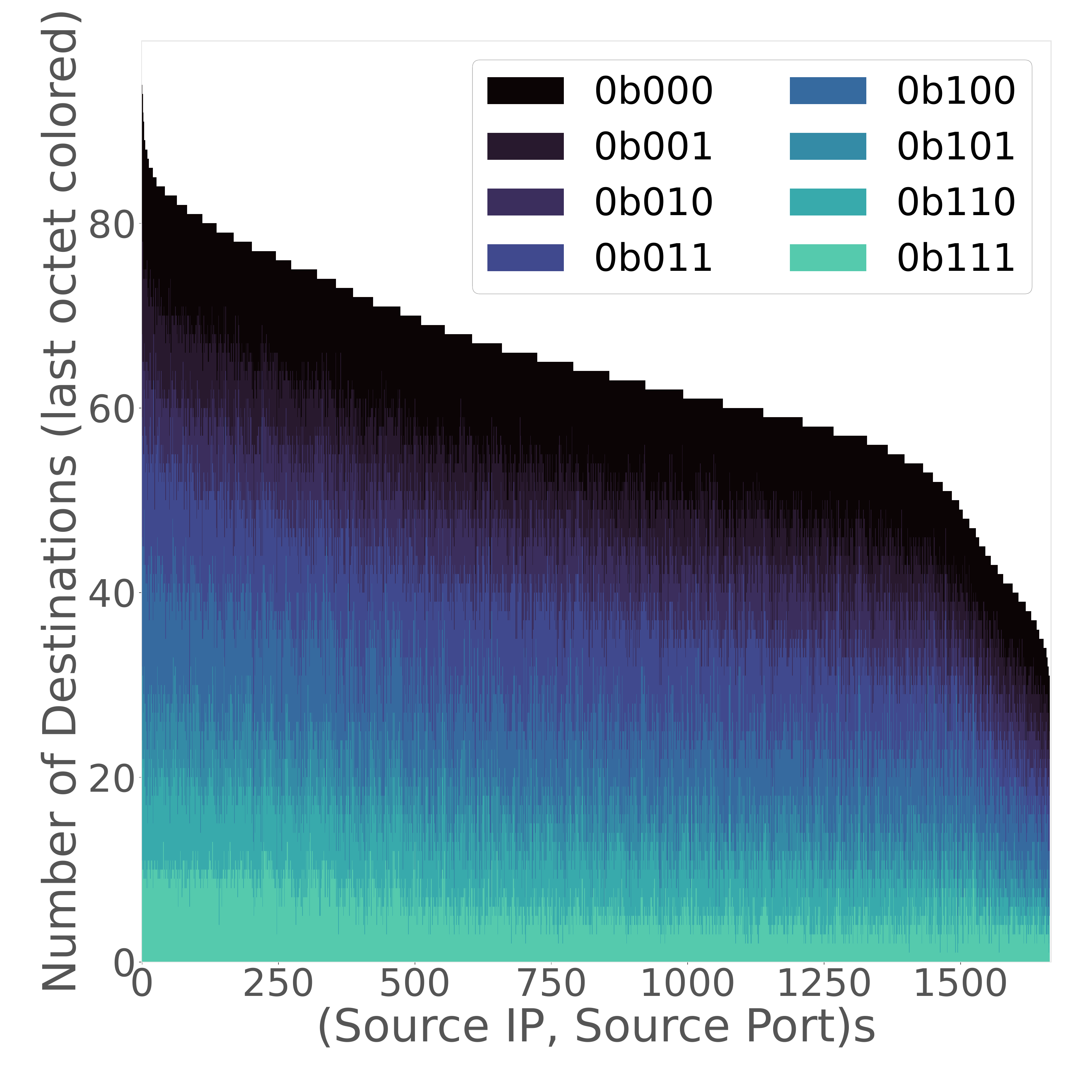}}\hfill
  \subfloat[Iran (\dns)\label{fig:ir-dns}]{\includegraphics[width=0.3\textwidth]{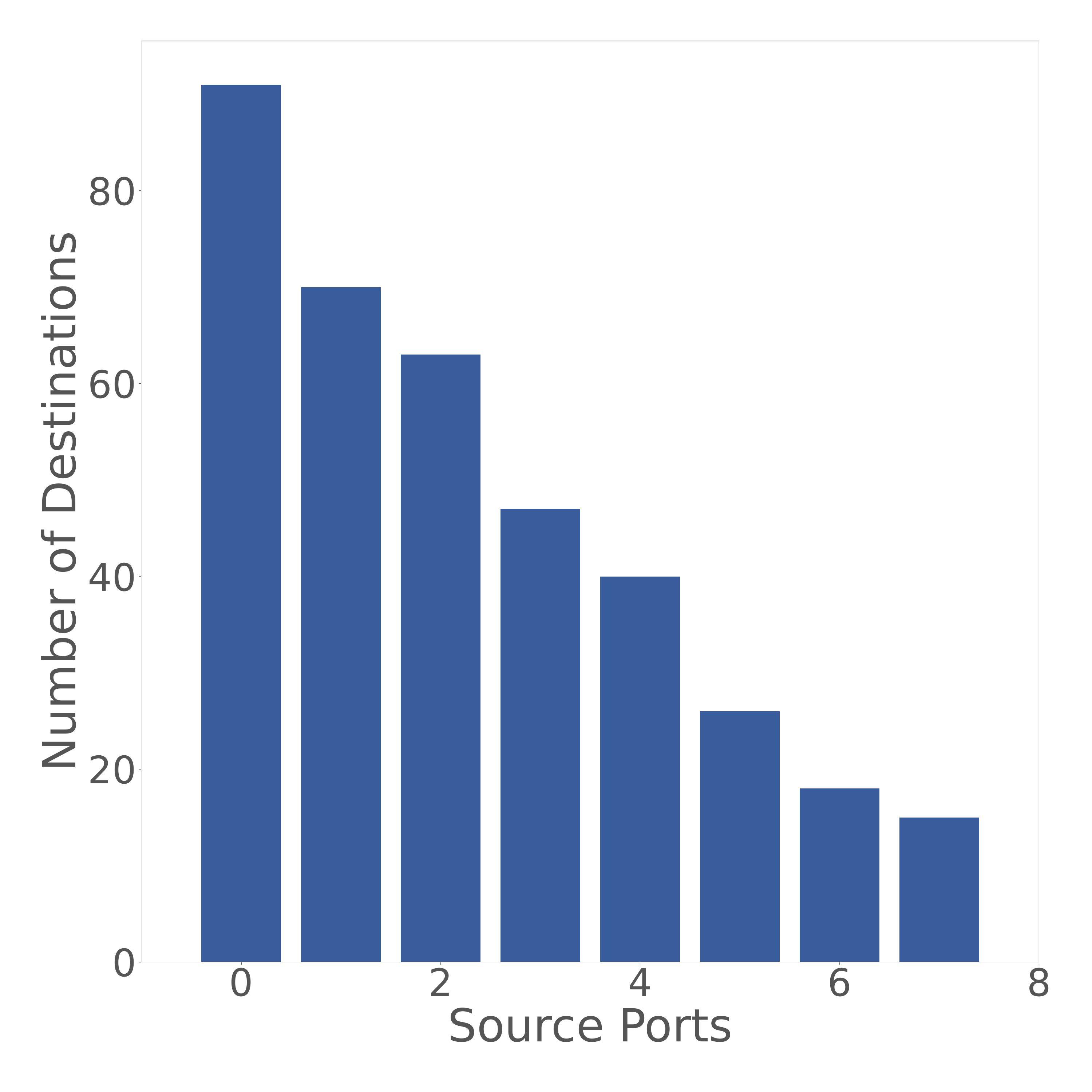}}
  
  \caption{\emph{Influence of Source IPs and/or Ports on Changes in
    Censorship.} Sources and destinations are sampled uniformly
    across the lowest 3 bits, and colored based on those bits for either source or destination IPs.
    X-Axis sorted.
  }
  \label{fig:bitPatterns}
\end{figure*}

Figure~\ref{fig:oddifacepcent} shows a CDF of (Source
IP, Source Port) combinations that demonstrate no censorship, across
varying destinations, broken out by country and protocol. We limit
this measurement to destinations that experience variation in
Table~\ref{tab:rq2}.  We observe that the per destination impact of
\ecmp routing varied broadly by country. With the percent of
(source IP, source Port) that observed no censorship for the median
\emph{affected destination} ranging from \medianAffectedSparamHigh in
the case of \medianAffectedHighWhere to as low as \medianAffectedSparamLow
in the case of \medianAffectedLowWhere. This calls for careful
selection of measurement parameters.

We also observe modal phenomena. For example, in ID (\http and \https) for
nearly $\sim$70\% of the destinations that are affected, $\sim$48\% of the (source
IP, port) combinations produced no censorship. We see similar patterns
in BD  with the split being at $\sim$40\% of the source parameters,
in BY with a split at around 10\%, in RU (\http) with a split at 50\%, and also patterns
in IN and RW. We speculate this could be routing algorithms that
divide traffic based on some fixed function of the packet, with some
paths lacking censorship.

We also observe that while for some countries the pattern of per-destination
impact follow similar trends for protocols within the country (\eg \http \&
\https with BD, BY, IN, \& ID), in others, the pattern \textit{varies even among
protocols} (\eg \http vs \https RU). It is interesting to note that while
\http \& \https follow similar trends for most countries, \dns
varies the most from the other two protocols. We speculate that this behavior
could be caused by either: 1) different censorship
infrastructure (possibly at different locations) for each of the protocols, or
2) the destination ports of these protocols themselves being used in load balancing~\cite{paris_traceroute}, yielding
different routes and changing censorship infrastructure.

\subsection{RQ3: \rqthree}
\label{sec:rq3}

We now seek to understand how particular
source IPs, source ports or the combination thereof impact censorship
variation. To achieve this we perform analysis on the low-order bit-patterns
(known to be used for routing~\cite{cisco_load_balance})
of both source and destination IPs compared to observed censorship variation.
Figure~\ref{fig:bitPatterns} shows a breakdown of five modal behaviors, discussed subsequently.
Results are colored by the lowest 3 bits of either source or
destination IP.

\begin{figure*}[]
  \centering
  \subfloat[ID (\https)\label{fig:id-https-graph}]{\includegraphics[width=0.8\textwidth]{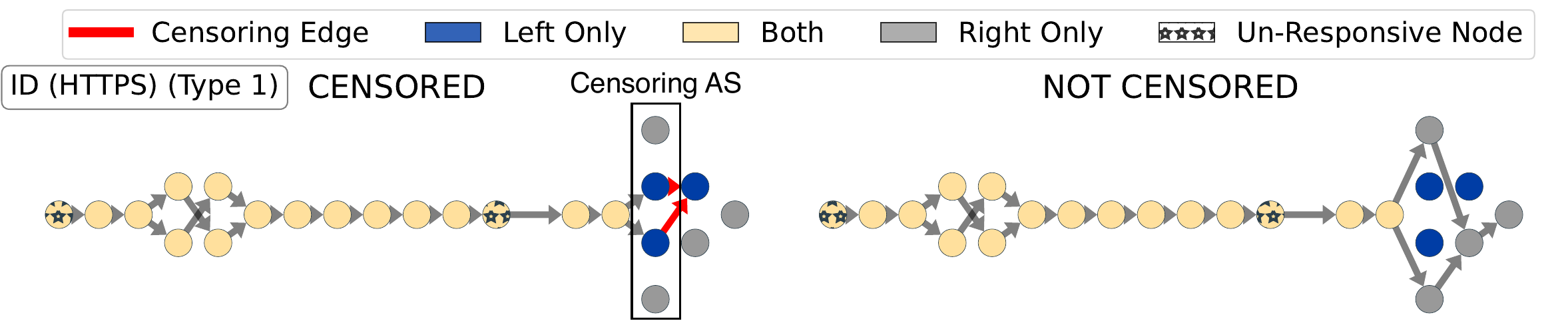}}\\
  \subfloat[IN (\http)\label{fig:in-http-graph}]{\includegraphics[width=0.8\textwidth]{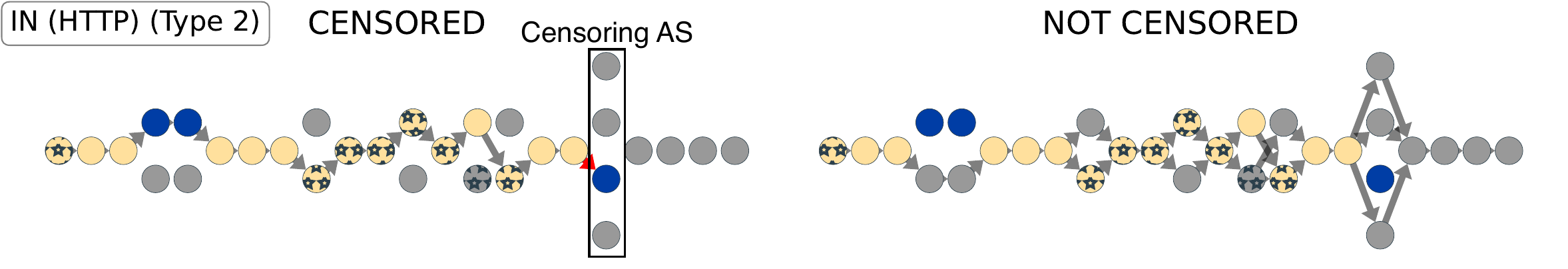}}\\
  \subfloat[KW (\http)\label{fig:kw-http-graph}]{\includegraphics[width=0.8\textwidth]{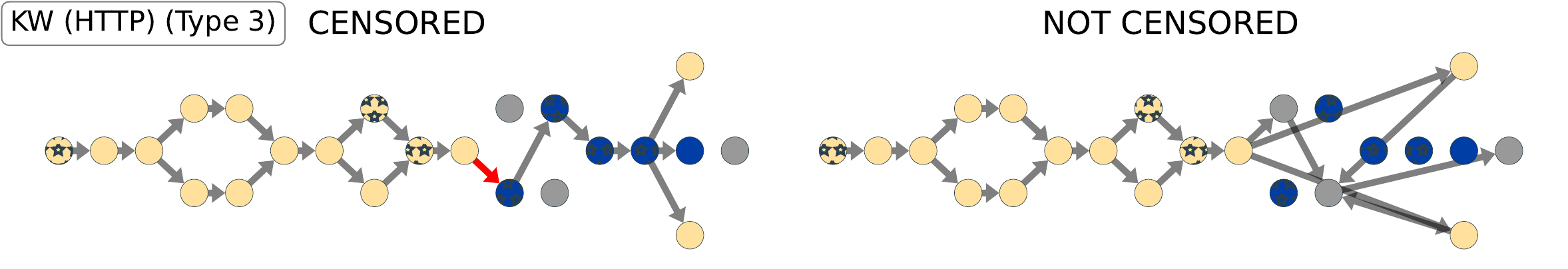}}\\
  \subfloat[PK (\http)\label{fig:pk-http-graph}]{\includegraphics[width=0.8\textwidth]{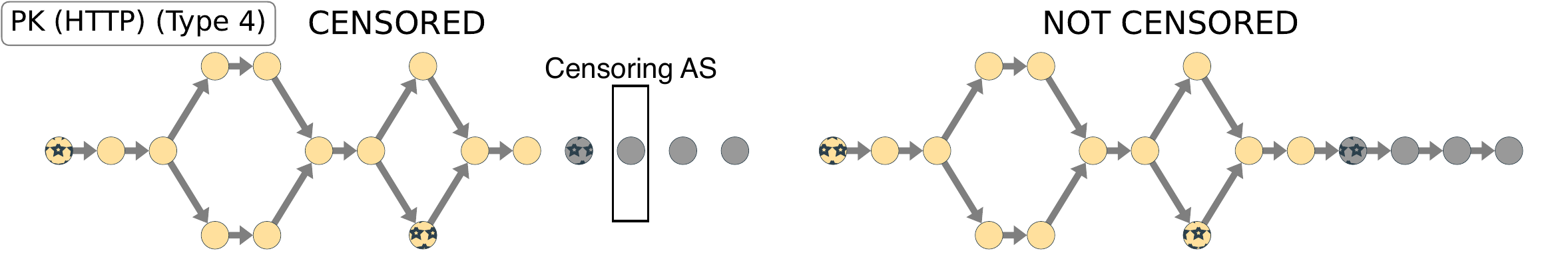}}
  
  \caption{
    Network graphs of censored routes (left) and
    non-censored routes (right). All nodes are shown in both, only routes change.
    Blue nodes were found only in paths that caused
    censorship, black found only in paths that had \textbf{no}
    censorship, yellow found in both, and censoring edges (if found) are red. 
    We observe notable structural differences between cases.
  }
  \label{fig:networkGraphs}
\end{figure*}

\PP{1. Particular source IPs see more censorship than others,
  associated with specific bit patterns in the source IP} 
Figure~\ref{fig:china-https} shows that based on the
source IP selected, the number of destinations that are impacted by
\ecmp routing more than double. It was interesting to note that: 1)
we only observed this particular effect with China and the Great
Firewall's censorship, and 2) the magnitude of lower order bits'
impact flipped in China (\https), compared to China
(\dns) from prior work~\cite{breadcrumb}. \ie 0b000 had the least
impact in \dns previously versus most impact in \https now.
This finding is consistent with and further supports the variation observed by
Anonymous~\cite{anon2014towards} where they found different injecting
interfaces simply by changing the source and destination of the probes
(with the GFW in China).

\PP{2. Particular source IPs have higher impact than others with
  no known associations with source IP bit patterns} In this case we
still see significant changes in the impact on particular source IPs
but cannot directly attribute them to any source IP bit
patterns. We see this particular effect in
Figures~\ref{fig:syria-http} and~\ref{fig:kr-http} where the impact
varies significantly by source IP, but each country denotes a distinct pattern with respect to destination IP bits. In addition we observe
modal patterns in Figure~\ref{fig:syria-http} which indicate
some source parameters are used differently for \ecmp routing,
changing censorship.

\PP{3. Uniform distribution of observed changes across the source IPs} In this
case the destination IPs yield uniformly distributed changes across all
source IPs tested, as seen in Figure~\ref{fig:in-http}. This likely indicates
hashing to perform \ecmp routing, vs specific bits.

\PP{4. Particular Source IP + Source Port combinations have higher impact
  than others} In this case a particular (source IP, source port)
combination produces significantly more variation than others. Figure~\ref{fig:bd-http} shows this, where the number of destinations impacted more than doubles between two (source IP, port)
combinations. This further points to the need to control both parameters.

\PP{5. Particular Source Ports have greater impact than others} In this
particular case, simply choosing different source ports has
significantly different impacts on the variation in censorship
measurement. We can observe this in Figure~\ref{fig:ir-dns} where the
number of destinations impacted by a particular source port can more
than double based on port selection. We note that: 1) we only observed this
behavior in the context Iran, and 2) this directly contrasts
prior work~\cite{breadcrumb} which did not observe any variation due to source port.

\subsection{RQ4: \rqfour}
\label{sec:rq4}

We have established that routing has a direct impact on the network paths of
packets and subsequently the results of common censorship measurement
techniques. In this section we seek to make qualitative associations between
differing censorship results and potentially deviating network paths as a
result of \ecmp routing in an attempt to understand the underlying effects that
cause variation in censorship. To achieve this we perform a deep dive to
understand the concrete causes for such variation across the different
countries and protocols. We find that \emph{such variation is not due to a
single global effect but rather a collection of different effects} ranging from
routing that appears to exercise failed/misconfigured censoring nodes to routing
that goes through completely different geographical regions that produce
different censorship \emph{en-route}.

We utilize \name to perform traceroutes and network graphs
(Section~\ref{subsec:traceroute}) for a geographically diverse set of
destinations for source parameters that showed variation, on a per
country/protocol basis. Choosing a diverse set of destinations allows
us to study potentially different effects at play even within a single
country. Using these network graphs we find \numtypes type of district
effects that contribute to \ecmp routing induced observed censorship
variation.

Figure~\ref{fig:networkGraphs} shows network graphs for a
representative set of countries and protocols that demonstrate the
different effects.  For each destination we group all experiments that
yield censorship into one graph (left) and all experiments that did
not yield censorship into another graph (right).
We now discuss these effects in greater detail in the context
of how they manifest in each of the different countries and protocols.
Appendix~\ref{sec:appendix_network_graphs} contains additional graphs.

\subsubsection{\typeone}

Here, the packet is routed from a single node at a particular hop to
a series of different nodes at the subsequent hop, that typically belong to an
AS known to perform censorship. These different nodes often exist in the same
/24 subnet and geographic region. \emph{We observe different censorship
behavior because some of these devices/paths perform censorship while others do
not}, pointing to potentially failed/misconfigured censoring devices. This was
one of the most commonly observed effects.

Identifying exact failed/misconfigured censoring nodes is
challenging without detailed knowledge of the underlying censoring infrastructure, but we can infer this from
comparing network graphs that have different observed censorship
behavior side-by-side. We do this by identifying particular nodes in
the censoring AS that are \emph{only} present in the paths (for source
parameters) that \emph{do not} exhibit censorship, suggesting
failure or misconfiguration.

\PP{Inter versus Intra AS} When routing to different nodes in the
censoring AS, \ecmp routing can take place either: 1) completely
within the censoring AS (Intra AS), \emph{or} 2) from a completely
different AS to the censoring AS (Inter AS). This highlights that
  variation in censorship measurement based on \ecmp is not solely due
  to routing differences inside censoring ASes, but
  also stems from routing changes yielding different ASes before
  the censoring AS.

We observe this effect in Indonesia (\https), shown in
Figure~\ref{fig:id-https-graph}, as the sensitive packet transits from
AS3491 (PCCW Global IP in Singapore) to AS7632 (PT Link Net, ISP in
Indonesia known to perform censorship~\cite{ooniindonesia}), it is
routed to several different IPs within several different /24 subnets
in AS7632, some of which experience (\rst) censorship while others do
not. We also note that these nodes 
are mutually exclusive for the two behaviors.

\begin{figure}[]
  \centering
  \includegraphics[width=\columnwidth]{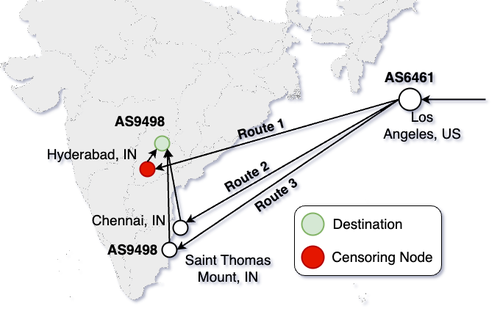}
  \caption{\ecmp routing yielding \emph{geographically} diverse
    routes. We observe that the sensitive packet is routed to
    different cities based on source IP or port, producing different
    observed censorship behavior. Censorship
    is only observed on Route 1. We note this exemplar is persistent.}
  \label{fig:india-type2}
\end{figure}

We also observe this behavior in: 1) Belarus (\http and
\https) as the packets traverse through a set of mutually exclusive
paths in AS6697 (Beltelecom, one of the ISPs known to perform
blockpage based censorship~\cite{oonibelarus, duncan2023detecting}),
where one transits nodes that always produce blockpage censorship and
the other passes through nodes with no censorship, 2) in Bangladesh
(\http) as the packet undergoes \ecmp routing from AS6939 (Hurricane
Electric IP in Singapore) or AS2914 (NTT America IP in
Singapore), to several different IPs in AS58717
(Summit Communications, a major ISP in Bangladesh known to perform
censorship~\cite{oonibangladesh}), some of which perform
blockpage/drop censorship, while others do not, 3) in Bangladesh
(HTTPS) where \ecmp routing entirely within AS17494 (BTCL, a large ISP
in Bangladesh) causes the sensitive \emph{client-hello} to pass
through certain nodes that never experience censorship while others
that always produce \rst censorship, 4) in Turkey (\https), as the
sensitive packet transits from AS1299 (Arelion Sweden IP in Germany)
to AS9121 (Turk Telekom, ISP in Turkey known to perform
censorship~\cite{jin2021understanding}), it is routed to different IPs
is the same /24 subnet in AS9121. On some of these paths, the
sensitive packet experiences censorship while on others it does not,
and 5) in South Korea (\http) where the packet transits from AS9848
(Sejong Telecom) to AS4670 (Shinbiro), the packet takes two distinct
paths going into AS4670; on one path the sensitive packet produces a
blockpage while on the other it passes through to the destination and
we get the response from the server. Although AS4670 is not known to
perform censorship, it has customers (\eg AS9848) that censor~\cite{hoang2019measuring}.

Such per-destination variation (presence or absence) of measured
censorship due to source parameters causing the packet to pass
through potentially failed/misconfigured censoring nodes could
possibly manifest as non-uniform censorship across different destinations
within a country/ISP, which could be a contributing factor in variation observation in prior work~\cite{iris, wang2017your,
  crandall2007conceptdoppler, weinberg2021chinese, Gill2015a,
  winters12foci, Ensafi2015a, raman2020measuring}.

\subsubsection{\typetwo}
\label{subsubsec:typetwo}

Here packets are routed from a particular
AS to a fixed endpoint, but along the way transit IPs that are in
different /24 subnets and are geographically diverse (but in the same
AS). As a result of such persistent substantial change in the
network path we observe different censorship behaviors.  Accounting
for such effects is critical for outside-in measurement so as to not
conflate route/path specific censorship behavior as censorship
behavior at the measurement destination.

We observe this in India (\http), shown in
Figure~\ref{fig:in-http-graph}, where the deviation in path producing
differing censorship behavior occurs when the packet transits
\textbf{\emph{from}} one of: 1) AS6461 (Zayo Bandwidth IP in London,
UK, or in Los Angeles, USA), or 2) AS7473 (Singapore Telecom in
Singapore)- \textbf{\emph{to}} AS9498 (BHARTI Airtel, known to perform
censorship~\cite{indianCensorship, singh2020india,
  katira2023censorwatch}). The IP in AS6461 (or AS7473) routes the
packet on IPs in noticeably different /24s in AS9498, sometimes in
completely different cities simply based on the packet
construction. Some of these paths produce a blockpage while others do
not. Only a handful of routers in AS6461 and AS7473
control the path taken into AS9498 and consequently
if we observe censorship. 

Figure~\ref{fig:india-type2} demonstrates this geographically diverse
\emph{censorship-along-the-way} effect, with packets between two fixed
endpoints taking different paths based on source parameters, those
paths varying geographically (per Ipinfo~\cite{ipinfo}), and
censorship correlating to geographic regions within India.  Only a
fraction of these paths pass through devices that perform censorship,
owing to larger impact of \ecmp on differences in observed censorship
in India (Table~\ref{tab:rq2}). Prior work found AS-level
variation in measured censorship for
India~\cite{katira2023censorwatch,india2018yadav} which highlights the
need to account for such \ecmp routing induced censorship changes, and
also the possibility that router regions, rather than end-host
regions, are being measured.

\subsubsection{\typethree}

In this mode we observe that at a particular hop, the
packet undergoes \ecmp routing, causing it to take completely
different AS paths before reaching converging at the destination. Some
of these paths never go through the censoring AS,
experiencing no censorship, while others do.

We observed this effect in Kuwait (\http, Figure~\ref{fig:kw-http-graph}), where depending on source parameters,
the packet either transits directly from AS3356 (LEVEL3 in London,
UK) to AS59605 (Zain Group), reaching the destination without censorship, \emph{or} it had unresponsive hops between AS3356 and AS59605,
where we observed censorship.

We also observed the effect in: 1) Kuwait (\https) as the packet
transits from a hop in AS3356 (LEVEL3, IP in UK), it either passes
through several unresponsive hops leading to an IP in AS47589 (Kuwait
Telecom, known to perform censorship~\cite{hoang2019measuring}), where
it terminates producing censorship, \emph{or} transits directly to
another IP in AS47589 eventually leading to the destination and no
censorship occurs, and 2) Rwanda (\http) as the packet transits from
AS16637 (MTN SA IP in Kenya) we observe two mutually exclusive paths,
one that transits to another AS16637 IP in SA then AS36890 (MTN
Rwandacell, large ISP in Rwanda) before reaching the destination,
another that passes through a set of non-responsive hops and reaches
the destination. The former path that passed through AS36890 always
experiences censorship, while the latter does not.

Such per-destination variation in censorship caused by simply changing
source parameters resulting in completely different AS paths not
experiencing censorship \emph{can also} manifest as non-uniform
country/ISP level, which could be a contributing factor in variation observation in prior work~\cite{iris,
  wang2017your, crandall2007conceptdoppler, weinberg2021chinese,
  Gill2015a, winters12foci, Ensafi2015a, raman2020measuring}.

\subsubsection{\typefour}

In this scenario the changes in censorship results cannot be directly
attributed to an observable change in path. This could occur due to different
forms censorship infrastructure in the path that do not answer with any \icmp
messages \emph{or} ones that are completely off-path.  We observed this
behavior with Pakistan (\http), shown in Figure~\ref{fig:pk-http-graph}, as the
packet transits from AS3356 (LEVEL3 IP in France) to AS17557 (Pakistan
Telecom, known to perform censorship~\cite{nabi2013anatomy, pakistanshutdown}),
there exists an unresponsive node where for some source parameters we observe censorship
(packet drops) and for others we do not.

We also observed this behavior in: 1) Iran (\http), where both sets
of source parameters have the same network path until they
reach an IP in AS58224 (Iran Telecom) at which point a blockpage is
always issued for certain set of source parameters while there is no
censorship for others, and 2) Algeria (\http and \https) where
\ecmp routing in AS174 (Cogent) produces mutually exclusive sub-paths
(each for a combination of source parameters) that exhibit different
observed censorship behavior. In this case as the packet leaves the
border router in AS174 towards AS36947 (Telecom Algeria), there are
several unresponsive hops before the destination.

\subsubsection{Result and Route Stability}
Our conservative selection criteria specifically controlled for stable
routes (by controlling for \flowid, Section~\ref{subsec:approach}).
Additionally, our experiments took place over a two week window,
requiring that the different routes 
and corresponding variation results were stable over that time horizon. We then
conducted a spot-check manual exploration approximately 2 months after the
initial experiments of the specific packet parameters and IPs that caused
variation, and found much of the variation remained. This is consistent with
prior work~\cite{paxson1997end}, that while dated, demonstrated that routes
tend to be stable over time.

\subsection{RQ5: \rqfive}
\label{sec:rq5}

\begin{figure}[]
  \centering
  \includegraphics[width=0.9\columnwidth]{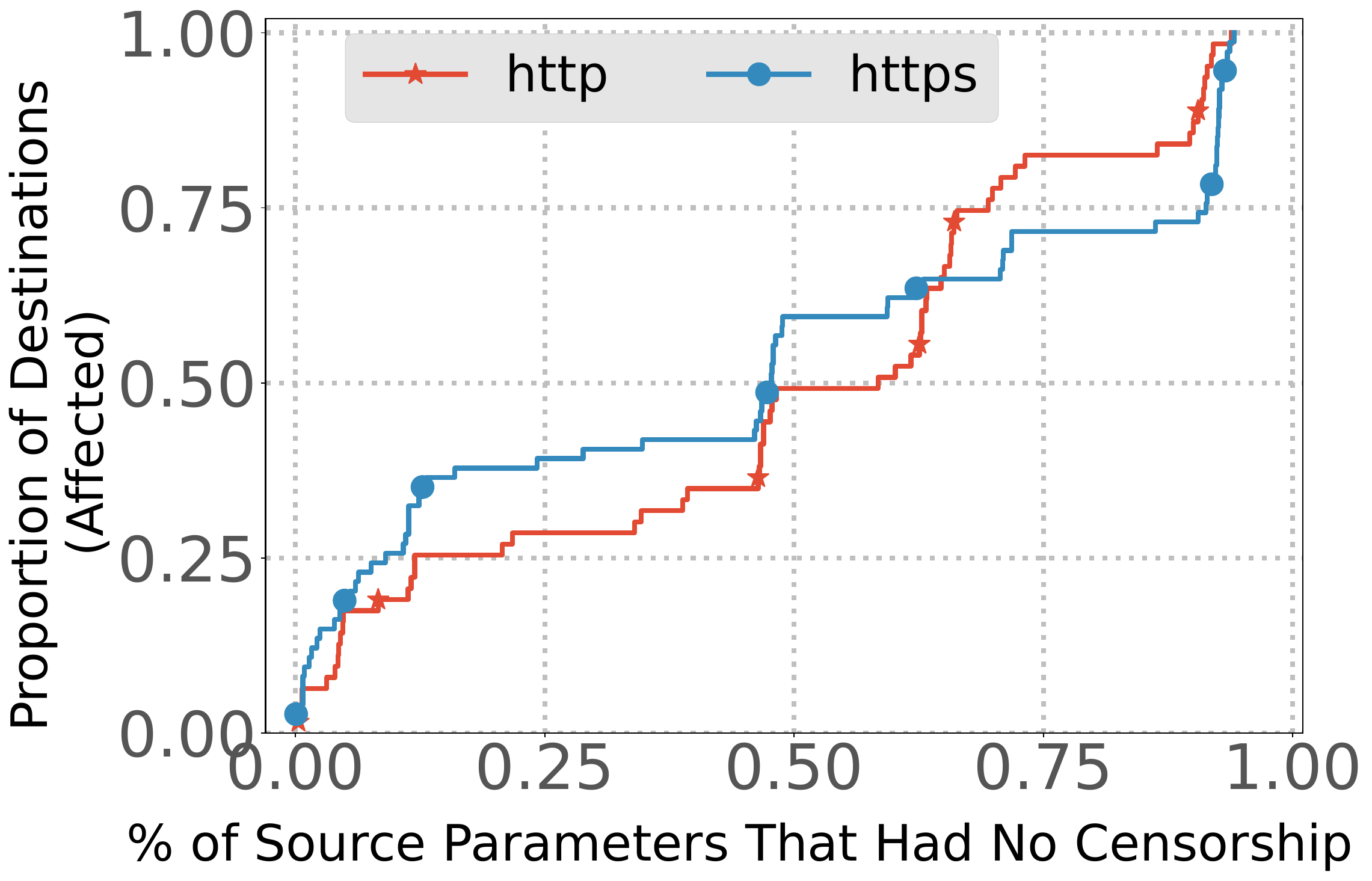}
  \caption{Simulation of impact of \ecmp on 
    a replication of prior work's~\cite{raman2020measuring} selection method. CDF of percent of (source IP, source Port) combinations per destination for
    which \emph{we} observed no censorship for the sample set of destinations.
    }
  \label{fig:oddiface-sample}
\end{figure}

We have qualitatively and quantitatively established the impact of
\ecmp routing on global outside-in censorship measurement while also
investigating the underlying causes of such variation. Finally we want
to contextualize prior results in the presence of \ecmp. We first look
at the applicability of \ecmp router induced censorship changes on the
different forms of variance observed in prior work (enumerated in
Section~\ref{subsec:prior_variance}). We then focus on specific
findings/results from prior work that are potentially impacted by
\ecmp routing~\cite{iris, raman2020measuring}.

\subsubsection{Applicability in Prior Work}

Section~\ref{subsec:prior_variance} detailed the different forms of
country and ISP level inconsistencies in measured censorship in prior
work. Not all of this variation is \ecmp routing induced, with the root
causes ranging from clearly demonstrated different censorship
implementations across a country~\cite{wright2014regional,
  xu2011internet, nisar2018incentivizing, india2018yadav,
  katira2023censorwatch}, to network load on censors manifesting as
sporadic absences in censorship~\cite{crandall2007conceptdoppler,
  Ensafi2015a}. We posit however that the per-destination changes in
censorship 
observed in our results 
by changing the source parameters of the sensitive packet, can manifest
as non-uniform censorship at a country/ISP
level as observed in prior work~\cite{iris, wang2017your,
  crandall2007conceptdoppler, weinberg2021chinese, Gill2015a,
  winters12foci, Ensafi2015a, raman2020measuring}. While we cannot
conclude that \ecmp routing is the (sole or contributing) cause for such
variation 
\ecmp routing is applicable to these methods and studies, and should
be accounted for to identify variation root-causes and stabilize measurements.
\ecmp routing is also applicable for external measurements
attempting to disambiguate ISP-granularity censorship differences, since prior
work has found concrete evidence of differing censorship
implementation at the ISP/AS-level~\cite{nisar2018incentivizing,
  india2018yadav, katira2023censorwatch} and our
findings (Section~\ref{subsubsec:typetwo}) suggest that for the same
destination, \ecmp routing causes the probe to pass through completely
different ASes with different censorship behavior.

\subsubsection{Potential Impact of ECMP routing on 2 Prior Works}

We now look at two prior outside-in studies focusing on: how their observations can
potentially be explained by \ecmp routing~\cite{iris}, and how their
results could be impacted by \ecmp routing~\cite{raman2020measuring}, 
based on our results.

\PP{Global DNS censorship measurement}
Our results can shed light on the causes of
heterogeneity in observed censorship seen by prior \dns censorship
measurement. \eg Figure 7 from Pearce~\et~\cite{iris} shows
banding effects at $\sim$10\% for Iran and $\sim$20\% in China,
which they attribute to non-determinism in censors.
When we compare their results to our own, we find that $\sim$7\% of destinations for
Iran, and $\sim$25\% of destinations for China are impacted by \ecmp routing.  While the
age of the study and lack of specific source parameters precludes definitive conclusion, the
similarity between results is consistent with \ecmp routing induced censorship
differences causing these banding effects. We note that Pearce~\et measures all forms of \dns manipulation, whereas we measure
only ``injected'' \dns manipulation, which limits our comparison.

\PP{Global HTTP and HTTPS censorship measurement}
In this case study we consider the example of a global outside-in
censorship study conducted by Raman~\et~\cite{raman2020measuring}. We
see that they choose either 11 (for \http) or 13 (for \https) destinations
per country for their study (from their Table
1~\cite{raman2020measuring}). To understand the potential impact of \ecmp
routing on such a study we randomly sampled 11 (for \http) or 13 (for \https)
destinations per country that observed censorship in our experiments for RQ2. We found
that across the countries out of 362 destinations that we sampled, \emph{138
($\sim$35\%)} destinations showed some form of variations in observed
censorship results. For this sample, Figure~\ref{fig:oddiface-sample}
shows the CDF of percent of source parameters that no censorship was
observed on per destination. We find that \emph{for a median affected
destination, $\sim$47\% of source parameters produced no censorship}. As such,
depending on the specifics of multiple experiments, repeated trials, and
specific source parameters used, prior work may have observed what appeared to
be failures or lack of censorship, when in reality they were observing
routing-induced censorship changes.

\section{Recommendations and Concluding Discussion}
\label{sec:lessons}
\label{sec:discussion}
\label{sec:conclusion}

Our work demonstrates the choice of intra-subnet source IP and ephemeral port
influences censorship measurement routes, and those routing changes in
turn impact observed censorship across \dns, \http, and \https, globally.
We show these variations are significant in terms of the number of
affected countries, IPs, and source parameters, and explore \emph{why} such changes exist.  We note that censorship measurement is a
critical tool for policymakers and evasion tool designers, and thus accurate
understanding of methods, results, and confounding factors such as those
discussed in this work are important for \emph{correct} assessment of information control. 

\PP{ECMP Routing Impacts Measurement, Globally}
Prior work observed the Chinese GFW's DNS injection changing based on
routing~\cite{breadcrumb}, but it was unclear if such behaviors were an
idiosyncrasy of that system or a global phenomenon rooted in routing.
We generalize our understanding of the intersection between routing and
censorship measurement, discovering that \ecmp routing's impact on censorship
measurement is both pervasive and significant.  We also observed differences in
how particular source parameters influence censorship results and found not
just source IPs but source ports can have an impact on censorship results.  These differences in changes point to the complexity of routing
and distributed systems, and call for further work in understanding localized
observed differences in censorship measurement globally. We further
contextualized results with prior work and found instances where
non-determinism in prior global censorship measurement is consistent with this
phenomenon. We also note that this phenomenon may impact other
forms of measurement, such as fast Internet scanning.

\PP{Causes of Variation.} We also explored \emph{why}
variation exists and found several causes: 1) \ecmp routing exercising
likely failed or misconfigured censoring devices, 2) \ecmp
routing causing a difference in AS path to cause ``routing around''
censorship behavior, and 3) \ecmp routing causing the packets to
traverse completely different geographical regions, producing different
censorship behavior \emph{en-route}. From this exploration we learn that in
addition to diversity in source parameters, censorship studies must account for
potentially observing different censorship behaviors \emph{along} the path
rather than \emph{at} the destination, when performing measurement.

\PP{Extending to IPv6 Censorship Measurement.}  IPv6 censorship is broadly under-studied,
irrespective of \ecmp routing.
Given that prior work~\cite{ipv6multipath} found that
\ecmp is \emph{more} prevalent (75\% of their measured routes) in IPv6 than IPv4,
we believe
that its impact on censorship measurement could be commiserate.  
Thus \ecmp routing should be taken into account as a first order concern as nascent
IPv6 censorship measurement studies are conducted. We note that the 
the presence of the \emph{Flow Label} field in IPv6
may aid in methods to produce route-stable censorship measurements.

\PP{Recommendations for Future Studies.}
We find that diversity in both source IP and source port is critical when
performing censorship measurement to avoid potentially incorrect
results or what appears to be transient failures, globally, across protocols.
We also note repetition is necessary to differentiate between
\emph{network effects} and \emph{actual censorship variation}. Measurements must be
repeated with significant parameter diversity, and localized
geographic effects must be validated with route diversity.
Finally, we find
significant differences in measured censorship in some countries by protocol.
These differences indicate a need for cross-protocol measurement
for a complete picture of Internet censorship.

\PP{Future Work and Evasion} Our work focuses on outside-in
measurement that is, by design, external.  Future work aimed at
extending these observations to volunteer vantage points within a
country is needed.  We however note such work is potentially
challenging as volunteer systems may not have access to numerous IPs,
or repeated experiments may be ethically challenging.  Further, the
prevalence across both countries and protocols, as well as the overlap
between protocols for given IPs, suggests the viability of future work
leveraging these differences to construct packets to
\emph{route-around} censorship, thus effectively evading. We note that
Tor Bridges~\cite{tor-bridges} are a particularly apt potential use.

\section{Acknowledgements}

The authors thank the anonymous reviewers and the shepherd for all
their thoughtful feedback provided during the review process. This
work was supported by funding from the National Science Foundation (NSF)
CAREER award 2239183.
 
\label{sec:acknowledgement}

\bibliographystyle{ACM-Reference-Format}
\bibliography{pearce_research,censor,misc,routing}

\newpage
\appendix

\section{Routing Path Variation}
\label{sec:appendix_variation}

\begin{figure}[]
  \centering \captionsetup{width=\columnwidth} \includegraphics[width=\columnwidth]{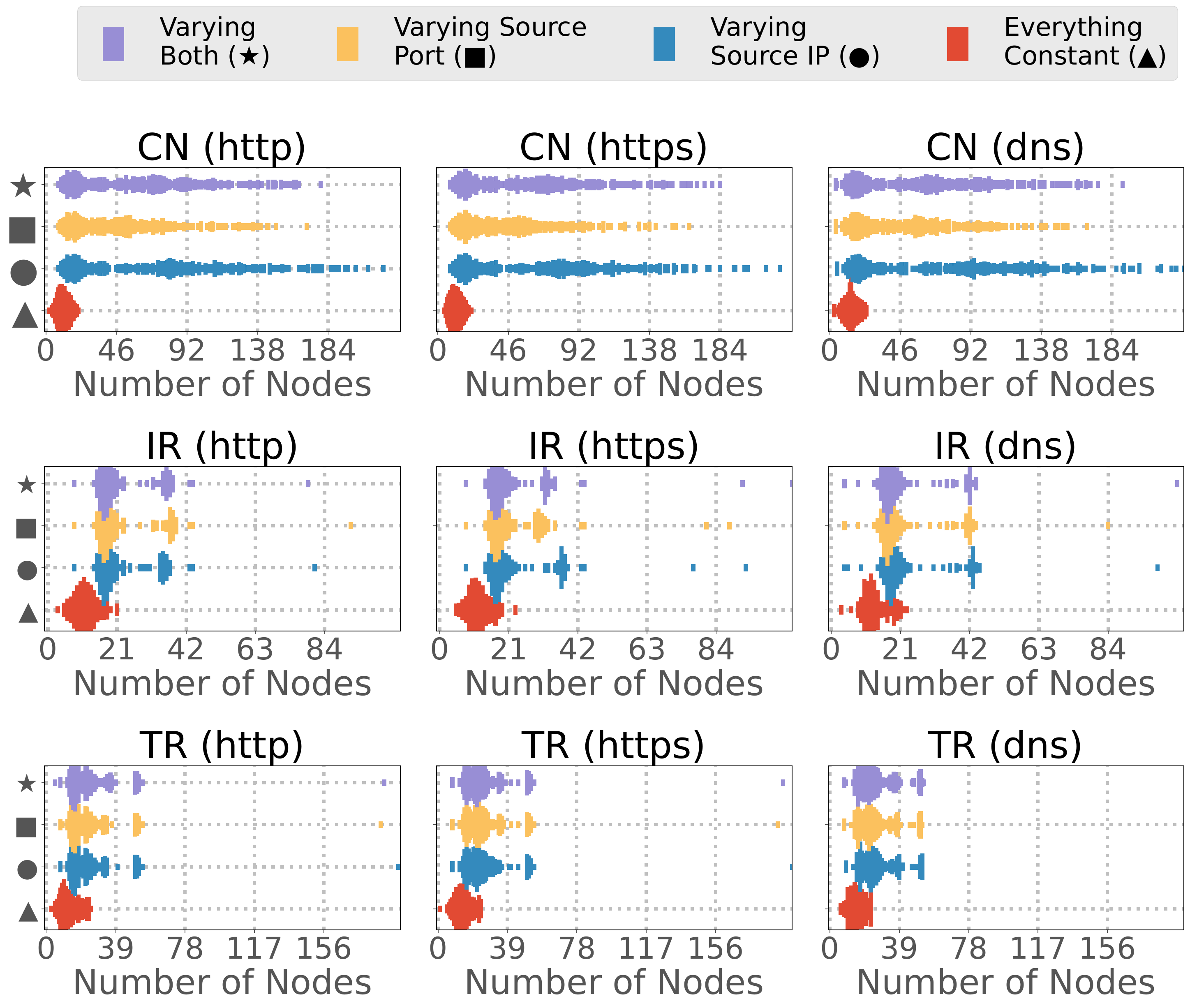} \caption{\emph{Normalized
  Distribution of Number Of Nodes across all destinations}. As
  with \emph{Number of paths}, observe that there are two modes of
  note: 1) varying source IPs has a greater impact than varying source
  port, and 2) Varying source IP and port have a comparable
  impact. We also note that for some countries (Iran),
  we see differences in the observed behavior \textit{among} different
  protocols.}  \label{fig:nodes_appendix}
\end{figure}

Figures~\ref{fig:nodes_appendix} shows
normalized distribution of \emph{Number of Nodes} (contrast to
Figure~\ref{fig:paths} which showed the normalized distribution
for \emph{Number of Paths}). We observe that
pattern of variations between protocols, for the same country remain quite
similar and variation in source port and IP produce various
modes.

\section{Overlap in Destinations Affected by \ecmp Routing
  Across Protocols}
\label{sec:appendix_overlap}
\begin{figure*}[]
  \centering
  \includegraphics[width=0.95\textwidth]{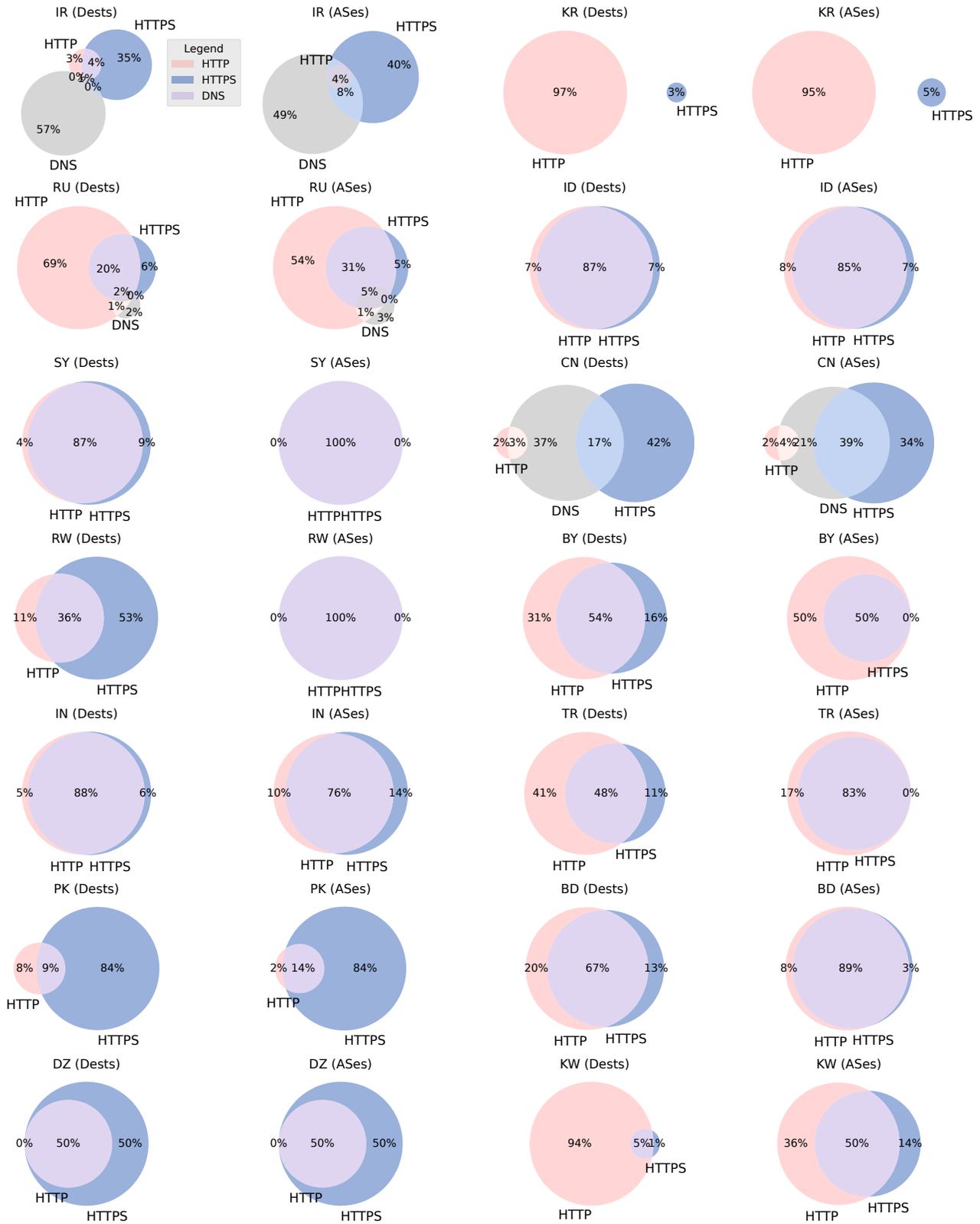}
  \caption{
    Venn diagram showing the overlap in destinations with censorship
    variation per protocol, by country. We find overlap across
    protocols varies significantly by country. 
    \label{fig:venn_variation}}   
\end{figure*}

We explore the overlap in the destinations (and ASes) impacted across various
protocols on a per-country basis. \ie do all destinations
experience variation equally across protocol, or does measurement of some
protocols vary with route more than others?

Figure~\ref{fig:venn_variation} shows with Venn diagrams for overlap
between protocols for all countries where we
observed censorship variation. This overlap between protocols varies for each
country with IN having the highest overlap at 88\% to KR with no
overlap. We note that in the case of RW even though
there is a 100\% overlap in ASes that are impacted, there is only
~35\% overlap in destinations that are impacted, potentially pointing
to different protocols being impacted by \ecmp routing differently. 

\section{RQ4 (cont'd): \rqfour}
\label{sec:appendix_network_graphs}
\begin{figure*}[]
\subfloat[BY (\http)]{\includegraphics[width=0.8\textwidth]{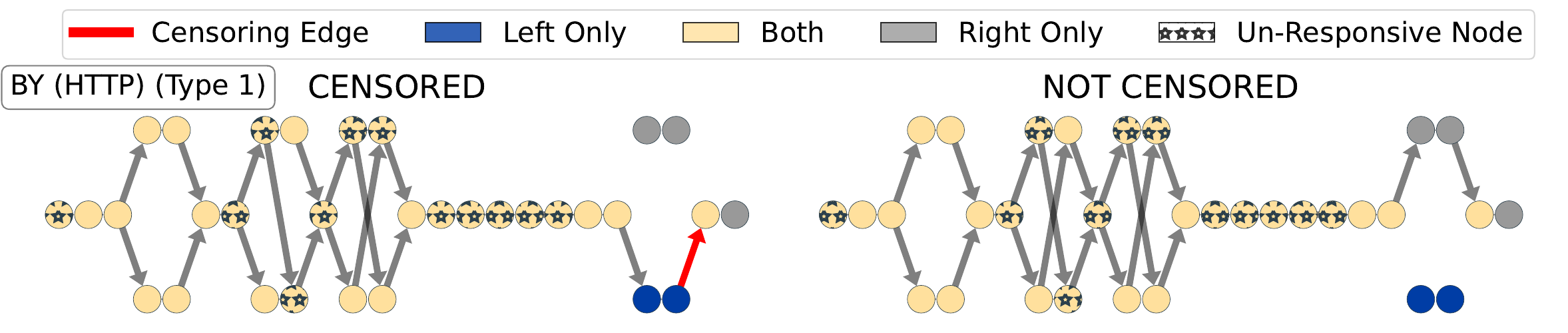}}\\
\subfloat[IN (\https)]{\includegraphics[width=0.8\textwidth]{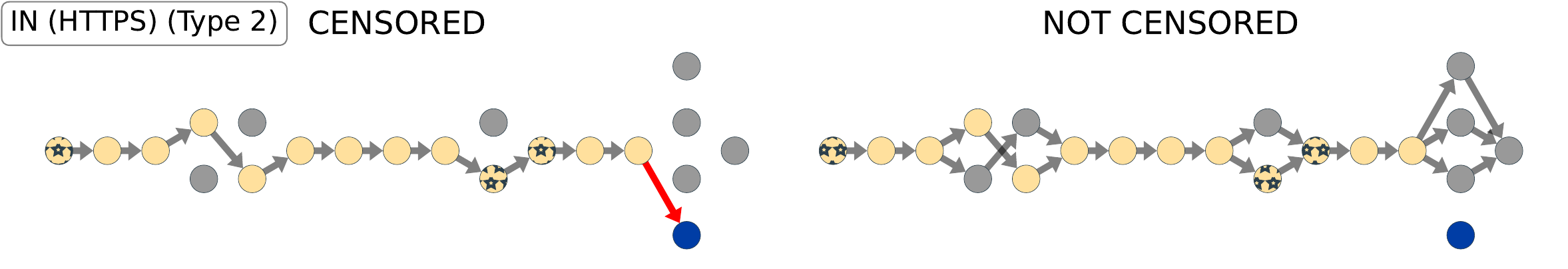}}\\
\subfloat[RW (\https)]{\includegraphics[width=0.8\textwidth]{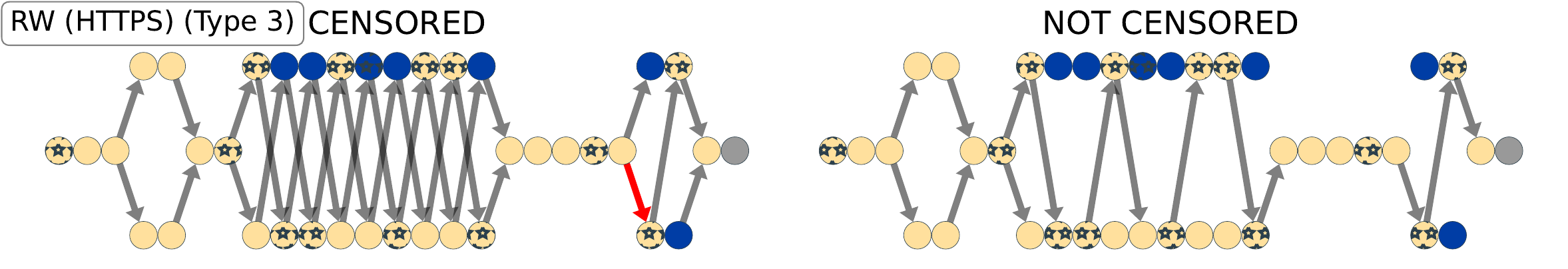}}\\
\subfloat[DZ (\http)]{\includegraphics[width=0.8\textwidth]{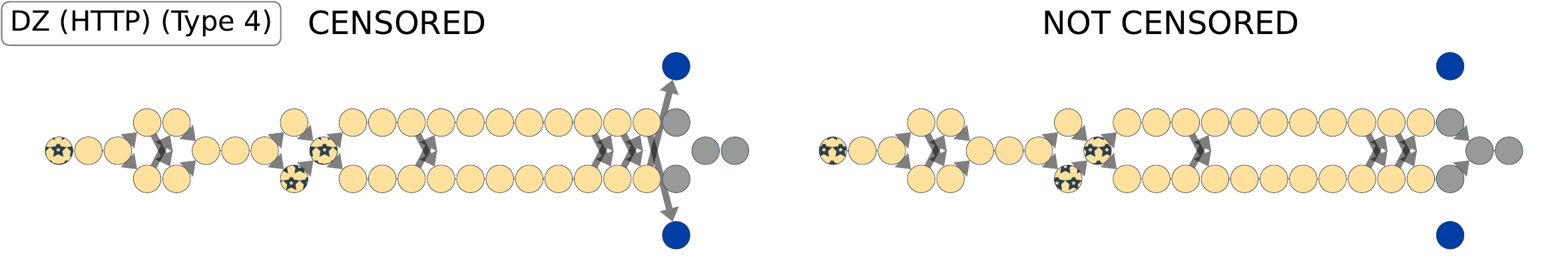}}\\
\subfloat[RU (\http)]{\includegraphics[width=0.8\textwidth]{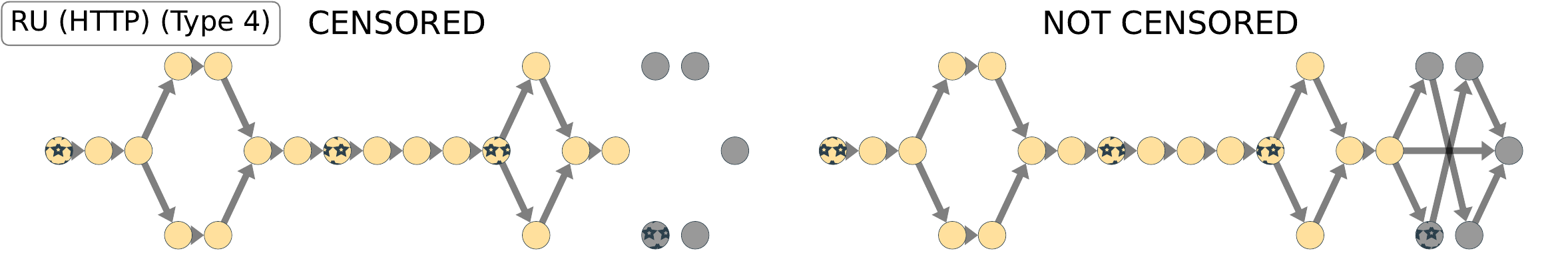}}\\
\subfloat[TM (\dns)]{\includegraphics[width=0.8\textwidth]{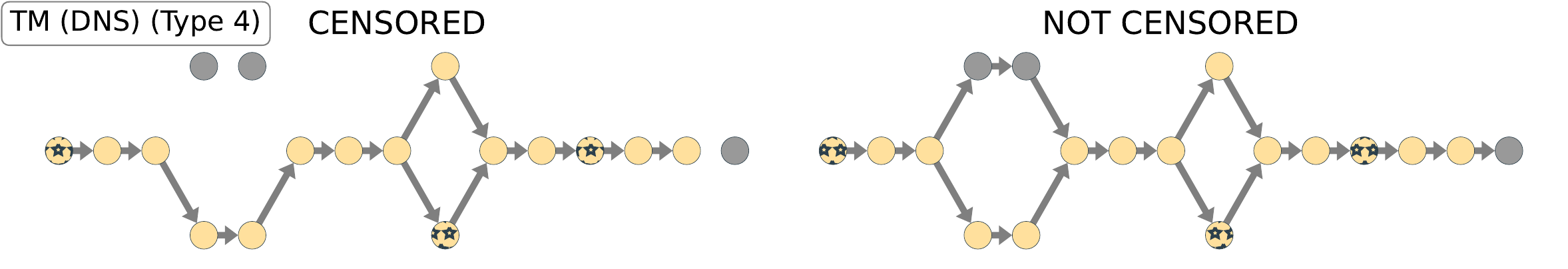}}
\caption{
Network graphs, extension of Figure~\ref{fig:networkGraphs} that demonstrates the effects described in Section~\ref{sec:rq4} for more countries.
}
\label{fig:networkGraphsAppendix}
\end{figure*}

Section~\ref{sec:rq4} explored the different underlying effects
causing \ecmp induced differences in censorship results. In this
section we present Figure~\ref{fig:networkGraphsAppendix}, which
provides more network graphs that showcase these effects in the other
countries.

\balance

\vspace{2em}

\begin{center}\textbf{Figures continue on next page.}\end{center}

\end{document}